\documentclass[10pt,journal,twocolumn]{IEEEtran}
\pdfoutput=1 
\usepackage{cite}
\usepackage{mathtools}
\usepackage{hyperref}
\usepackage{graphicx}

\usepackage{enumitem}
\usepackage{tikz}
\usepackage{bm}
\usepackage{amsmath}
\usepackage{algorithm}
\usepackage{algorithmic}
\usepackage{amsmath,amssymb,amsthm}
\usepackage{array}
\usepackage{relsize}
\usepackage{subcaption}
\usepackage{mathtools, nccmath}
\usepackage[labelformat=simple]{subcaption}
\usepackage[normalem]{ulem}
\newcolumntype{P}[1]{>{\centering\arraybackslash}p{#1}}
\DeclareMathOperator{\C}{\mathbb{C}}
\DeclareMathOperator{\E}{\mathbb{E}}
\DeclareMathOperator{\R}{\mathbb{R}}
\DeclareMathOperator{\N}{\mathcal{N}}

\DeclareMathOperator{\CG}{\mathcal{C}}

\DeclareMathOperator*{\argmax}{arg\,max}
\DeclareMathOperator*{\argmin}{arg\,min}
\DeclarePairedDelimiter\abs{\lvert}{\rvert}
\DeclareCaptionSubType * [alph]{table}
\newcommand{\x}{\mathbf{X}}
\newcommand{\A}{\mathbf{A}}
\newcommand{\B}{\mathbf{B}}

\newcommand{\h}{\mathbf{H}}
\newcommand{\f}{\mathbf{F}}

\newcommand{\I}{\mathbf{I}}
\newcommand{\z}{\mathbf{Z}}

\newcommand{\U}{\mathbf{U}}

\newcommand{\fs}{\mathbf{f}}
\newcommand{\rt}{\mathbf{R}}

\newcommand{\hs}{\mathbf{h}}

\newcommand{\xs}{\mathbf{x}}

\newcommand{\xxn}{\bm{\mathsf{x}}}

\newcommand{\ysn}{\mathsf{y}}

\newcommand{\ssn}{\bm{\mathsf{s}}}
\newcommand{\ssf}{\mathsf{s}}

\newcommand{\ys}{\bm{\mathsf{y}}}
\newcommand{\ws}{\bm{\mathsf{w}}}

\newcommand{\wsn}{\mathsf{w}}

\newcommand{\hr}{\mathsf{H}}
\newcommand{\tr}{\mathsf{T}}

\newcommand{\bc}{\begin{center}}
\newcommand{\ec}{\end{center}}
\newcommand{\norm}[1]{\left\lVert#1\right\rVert}
\newcommand{\ds}{\displaystyle}
\newcommand{\uprightsubscript}[1]{_{\textnormal{#1}}}
\begingroup\lccode`~=`?\lowercase{\endgroup\let~}\uprightsubscript
\AtBeginDocument{\mathcode`?="8000 }

\newcommand{\tn}[1]{\textnormal{#1}}

\theoremstyle{remark}

\definecolor{dg}{RGB}{255,0,0}
\graphicspath{ {./images/} }

\begin{document}

\title{Lens-Type Redirective Intelligent Surfaces for Multi-User MIMO Communication}
\author{\IEEEauthorblockN{{Bamelak Tadele}, Faouzi Bellili, \textit{Member, IEEE}, Amine Mezghani, \textit{Member, IEEE}, {Md Jawwad Chowdhury, and Haseeb Ur Rehman, }} 
\thanks{The authors are with the Department of Electrical and Computer Engineering at the University of Manitoba, Canada (emails: \{{tadeleb}, chowdhmj, urrehmah\}@myumanitoba.ca, \{Faouzi.Bellili, Amine.Mezghani\}@umanitoba.ca). Work supported by {the Discovery Grants Program of}
the Natural Sciences and Engineering Research Council of Canada (NSERC) and Futurewei Technologies Inc.}}
\date{}
\maketitle

\begin{abstract}
This paper {explores} the idea  of using \textit{redirective} reconfigurable intelligent surfaces (RedRIS) to overcome many of the challenges associated with the conventional \textit{reflective} RIS. We develop a framework for jointly optimizing the switching matrix of the lens-type RedRIS ports along with the active precoding matrix at the base station (BS) and the receive scaling factor. A joint non-convex optimization problem is formulated under the minimum mean-square error (MMSE) criterion with the aim to maximize the spectral efficiency of each user. In the single-cell scenario, the optimum active precoding matrix at the multi-antenna BS and the receive scaling factor are found in closed-form by applying Lagrange optimization, while the optimal switching matrix {of} the lens-type RedRIS is obtained by means of a newly developed alternating optimization algorithm. We then extend the framework to the multi-cell scenario with single-antenna base stations that are aided by the same lens-type RedRIS. We further present two methods for reducing the number of effective connections of the RedRIS ports that result in appreciable overhead savings while enhancing the robustness of the system. The proposed  RedRIS-based schemes are gauged against {conventional} reflective RIS-{aided} systems under both perfect and imperfect channel state information (CSI). The simulation results show the superiority of the proposed schemes in terms of overall throughput {while incurring} much less control overhead.

\end{abstract}


\section{Introduction}

\subsection{Background}

Wireless communication technology is developing at a fast pace to meet the ever-growing demands for higher data rates and massive connectivity. The soaring number of users in the network is exacerbating the interference problem which limits the performance of legacy wireless networks. To resolve the problem of spectrum shortage, the wireless technology is migrating to higher frequency bands to leverage the under-utilized spectra \cite{9086766}. With the fifth-generation (5G) mmWave communication technology being currently available to a limited demography, researchers have already started exploring the feasibility of using the Terahertz (THz) band in beyond 5G (B5G) networks \cite{9326394,9690477}. However, THz communication faces many challenges such as severe wireless propagation loss and difficulties related to transceiver design \cite{8387211,9618776}. Recently, the reconfigurable intelligent surface (RIS) technology has emerged as a cost-effective solution for purposely controlling the radio propagation environment to extend the wireless coverage and better mitigate the interference issues \cite{8647621,8917871,8811733,renzo2019smart}. The widely studied reflective RIS consists of a planar metasurface with several passive reflective elements which alter the incident signal by means of adaptively {controlled} phase shifters. {In this design, each element in the RIS locally reflects the incident signal independently of the other elements leading to a diagonal phase-shift matrix. Significant research effort has been made towards the design of the reflective RIS for passive beamforming purposes in single-user \cite{9362274,9473585,8930608,9198125,8683145,9206080,9115725,9226616,8741198,8796365} and multi-user \cite{wu2019intelligent,wu2019beamforming,abeywickrama2020intelligent,pan2020multicell,vampmain,wijekoon2024phase} MIMO communication.} {A comprehensive survey of reflective RIS can be found in \cite{liu2021reconfigurable}.} 

{While these ideal reflective RIS models based on local designs have demonstrated performance improvements, a careful analysis shows that passive lossless local surfaces cannot create arbitrary wavefront transformations without producing parasitic beams in undesired directions \cite{shastri2023nonlocal}. 
If the lossless constraint is relaxed and power can be absorbed locally, perfect reflection is possible at the cost of degraded efficiency at large angles. Achieving ideal lossless reflection requires a strongly non-local response where power may be absorbed or generated locally while being conserved over the entire surface \cite{asadchy2016perfect,estakhri2016wave}. Under this paradigm, the load scattering matrix (analogous to the phase shift matrix) becomes non-diagonal and is known in the literature as non-local redirective RIS (RedRIS) \cite{mezghani2022reconfigurable} or beyond diagonal RIS (BD-RIS) \cite{shen2021modeling}. A non-diagonal design also provides more flexible beam management as the RIS elements are allowed to cooperate with one another. There has been recent progress made on the BD-RIS configuration \cite{shen2021modeling,li2022reconfigurable,nerini2023discrete,nerini2023closed,nerini2024beyond}. The proposed designs are based on interconnecting the RIS elements by tunable impedance components and range from the disconnected (i.e., diagonal) RIS to a fully-connected RIS where all the elements are interconnected. The major drawback of these designs is that the gained performance improvements occur at the cost of additional circuit complexity and control overhead. Indeed, this has led to the study of the Pareto frontier for the performance-complexity tradeoff \cite{nerini2023pareto}. While prior work on BD-RIS has thoroughly studied non-diagonal structures, there has been limited attempts to exploit the channel properties so as to decrease the complexity of the non-diagonal designs.}

In this paper, we {investigate} the idea of using a lens-type redirective RIS
whose back-to-back port connections can be appropriately configured to select the beamforming vectors that redirect the incident signals towards their intended users\footnote{{An in-depth study of redirective RIS can be found in \cite{mezghani2022nonlocal}. Additionally, the network controlled repeater \cite{ayoubi2022network}, a standardized 5G component, can also be regarded as a particular type of RedRIS combined with an amplifier within the port interconnection.}}. Such beamforming vectors are selected from a DFT codebook that scans all the directions of radiation \cite{zhang2021learning}. We propose a general framework which involves the optimization of the back-to-back connections (or switching matrix) between the RedRIS ports, as opposed to finding the optimal phase shifts in the purely reflective RIS. {In doing so, we exploit the angular-domain sparsity of mmWave/THz channels owing to the inherent DFT transform  performed by the lens on the incident waves. This appreciably reduces the incurred signaling overhead and operational complexity for RIS-aided systems. In fact, the control overhead scales only logarithmically with the RedRIS size \cite{mezghani2022nonlocal} in contrast to the reflective RIS where the control overhead scales linearly with the number of phase shifters \cite{https://doi.org/10.48550/arxiv.2003.02538,9039554} and even more so in BD-RIS systems.} Although the lens-type RedRIS is a passive network element, it can be integrated with two-port amplifiers contrarily to the reflective RIS which would require reflection-mode amplifiers where stability becomes a major issue \cite{6648436,mezghani2022nonlocal}. To the best of our knowledge, {aside from our previous work in \cite{mezghani2022nonlocal,mezghani2022reconfigurable}}, the use of lens-type RedRIS for electromagnetic radiation control has not been investigated in the open literature. {Such} lens-type smart surface {is called} redirective RIS (RedRIS) due to its capabilities of wirelessly redirecting the impinging waves to their intended destinations in the downlink and uplink modes alike. We use the names RedRIS and lens-type RIS interchangeably throughout the paper.


\subsection{Contributions}
{We consider a lens-type RedRIS-aided wireless communication system serving multiple users in the downlink for both the single- and multi-cell settings. The lens-type RedRIS has a large number of antenna elements each of which with its dedicated port. Control of the RedRIS is {done} through back-to-back port switching connections which are optimized so as to wirelessly route the incoming signals {towards} their intended users. We propose a dynamic solution that optimizes the switching matrix of the RedRIS ports for the single- and multi-cell scenarios. For the single-cell setup, we also optimize the active precoding matrix at the multi-antenna BS together with the receive scaling factor, while for the multi-cell setup with single-antenna BSs we only optimize the different receive scaling factors. The main contributions of this paper are as follows:

\begin{itemize}
    \item First, we consider the single-cell scenario and formulate a joint optimization problem for the design of the switching matrix at the lens-type RedRIS along with the BS precoding matrix and the receive scaling factor under the sum MMSE criterion. We do so by minimizing the received symbols error of all users thereby maximizing the overall spectral efficiency.
    \item We apply alternate optimization \cite{dimitrinl} to break the overall optimization problem into two sub-optimization tasks. One sub-problem consists in optimizing the BS precoding matrix and the receive scaling factor, while the other deals with the optimization of the port switching matrix at the RedRIS.
    \item We find the optimum BS precoding matrix and the receive scaling factor in closed-form expressions using Lagrange optimization.
    \item To optimize the switching matrix, we develop an alternating optimization algorithm which consists of two modules: $i)$ regularized least-squares (R-LS) module and $ii)$ projection module.
    \item We also develop two methods to reduce the number of back-to-back connections between the RedRIS ports. In the first method, a small number of effective switches are selected for the multi-user scenario based on each user's MSE criterion. The second method is for the single-user scenario where the incident signal is redirected by connecting two RedRIS ports only. Such reduction in the number of port connections results in remarkable computational and control overhead savings compared to the fully-connected configuration. 
    \item We extend the framework to the multi-cell case wherein each cell is equipped with a single-antenna BS. We jointly optimize the switching matrix and the different receive scaling factors by means of the proposed algorithm.
    \item We gauge the performance of the proposed schemes against the widely-studied reflective RIS-aided communication. The simulation results reveal that using a lens-type RedRIS with a limited number of effective port connections yields considerably higher throughput for both single- and multi-cell scenarios. Moreover, it will be seen that reducing the number of RedRIS back-to-back connections incurs a slight performance loss provided that the effective ports are appropriately selected as explained later on in this paper. Finally, we test the robustness of the proposed solution by assessing its performance under residual channel estimation errors (i.e., imperfect CSI knowledge).
\end{itemize}
}
\raggedbottom
\subsection{Paper organization and notations}
The remainder of the paper is organized as follows. The system model along with the assumptions and problem formulation are described in Section \hyperref[sec:systemmodel]{II}. In Section \hyperref[sec:algorithm]{III}, we develop an algorithm for the joint optimization of the BS active precoding matrix and the receive scaling factor together with the ports switching matrix at the RedRIS. In Section \hyperref[sec:solution]{IV}, the proposed algorithm is used to solve the joint optimization problem for the single-cell scenario and Section \hyperref[sec:reduction]{V} discusses two methods for reducing the number of RedRIS port connections for the single- and multi-user cases. Extension of the proposed framework to the multi-cell scenario is provided in Section \hyperref[sec:multiple-BS]{VI}. Simulation results are provided in Section \hyperref[sec:numeric-res]{VII} to illustrate the performance of the proposed RedRIS-based scheme, before concluding the paper in Section \hyperref[sec:conclusion]{VIII}.

\vspace{0.1cm}

\noindent\textbf{Notations}: In this paper, scalar variables are represented by lower-case italic letters (e.g., $x$). Lower- and upper-case bold fonts, $\xs$ and $\x$, are used to denote vectors and matrices, respectively. The $(m,n)$th entry of $\x$ is denoted as $\x_{mn}$, and the $n$th element of $\xs$ is denoted as $x_n$.  $\tn{Rank}(\mathbf{X})$ and $\tn{Tr}(\mathbf{X})$, return the \tn{rank} and the trace of $\mathbf{X}$, respectively. The identity, all-ones, and all-zero matrices of size $N \times N$ are denoted as $\I_N$, $\mathbf{J}_N$, and $\mathbf{0}_{N}$, respectively. Moreover, $\tn{Diag}(\mathbf{X})$ sets the off-diagonal elements of $\mathbf{X}$ to zero and $\textrm{Diag}(\mathbf{x})$ returns a diagonal matrix whose main diagonal is $\mathbf{x}$. The shorthand notation $\xxn \sim \CG\!\N(\xs;\mathbf{m},\rt)$ means that the random vector $\xxn$ follows a complex circular Gaussian distribution with mean $\mathbf{m}$ and covariance matrix $\rt$. Moreover, $(.)^*$, $\ (.)^\tr$, and $\ (.)^\hr$ stand for the conjugate, transpose, and Hermitian (transpose conjugate) operators, respectively. We use $\tn{vec}(.)$ and $\tn{unvec}(.)$ to denote vectorization of a matrix and unvectorization of a vector back to its original matrix form, respectively. In addition, $\vert.\vert$, $\|.\|?2$ and $\|.\|?F $ stand for the modulus, Euclidean norm, and Frobenius norm, respectively. The cardinal of  a set $\mathcal{E}$ is denoted as $|\mathcal{E}|$ and $\Re\{.\}$ stands for the real part of its complex-valued argument. The statistical expectation is denoted as $\E\lbrace.\rbrace$, and the operator $\otimes$ denotes the Kronecker product between two matrices.

\section{System model, assumptions, and problem formulation}
\label{sec:systemmodel}
\begin{figure}
\bc
\scalebox{0.85}{
\begin{picture}(290,180)
\put(0,0){\includegraphics[scale=0.45]{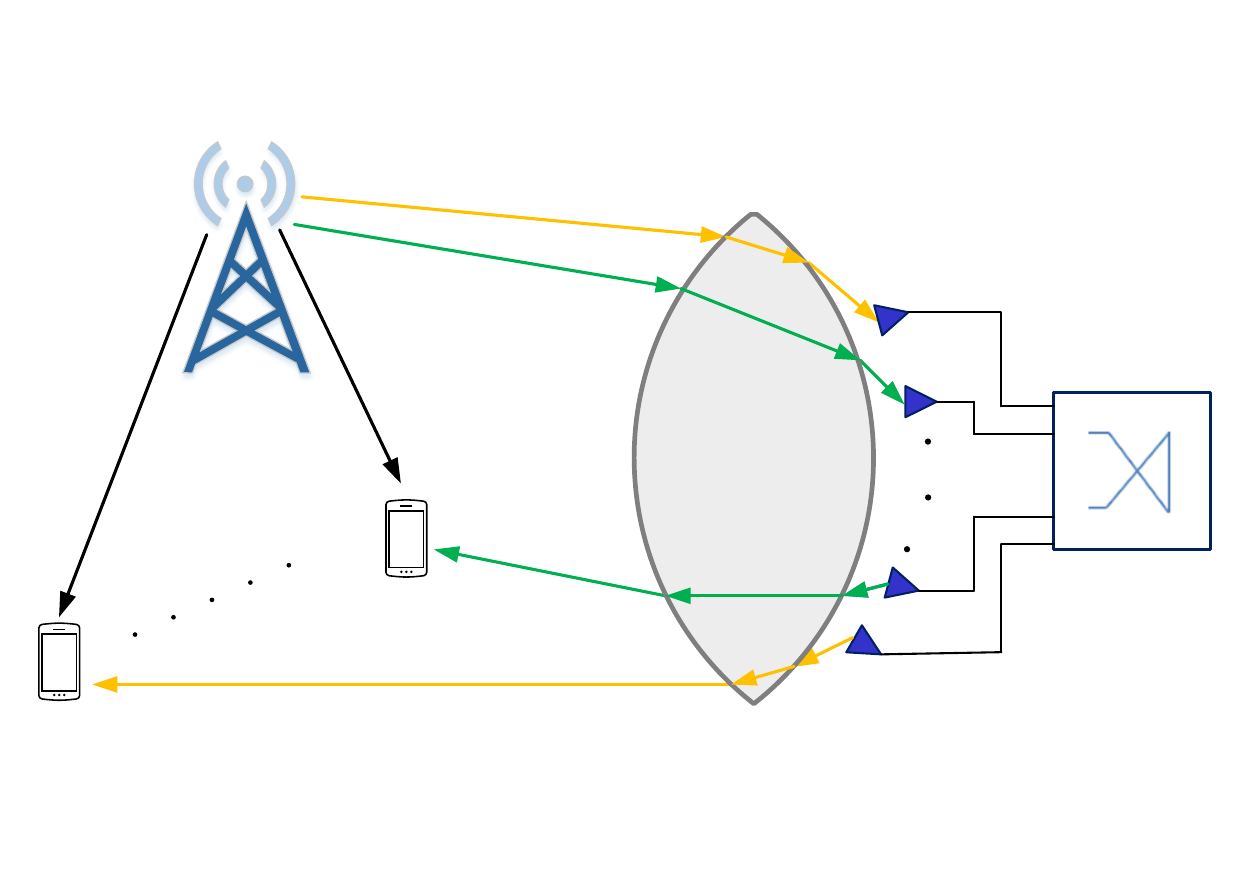}}
\put(46,164){BS}
\put(2,25){\footnotesize{User $M$}}
\put(77,53){\footnotesize{User $1$}}
\put(150,95){\footnotesize{RedRIS}}
\put(151,85){\footnotesize{with $K$}}
\put(155,75){\footnotesize{ports}}
\put(272,20){\rotatebox{90}{\footnotesize{back-to-back connection of RedRIS ports}}}
\put(105,147){$\h?{b-s}$}
\put(110,56){$\hs_{\tn{s-u},1}$}
\put(110,30){$\hs_{\tn{s-u},M}$}
\put(83,98){$\hs_{\tn{b-u},1}$}
\put(27,77){$\hs_{\tn{b-u},M}$}
\end{picture}
}
\caption{RedRIS-assisted multi-user MIMO system.}
\label{fig:sys-model}
\ec
\end{figure}
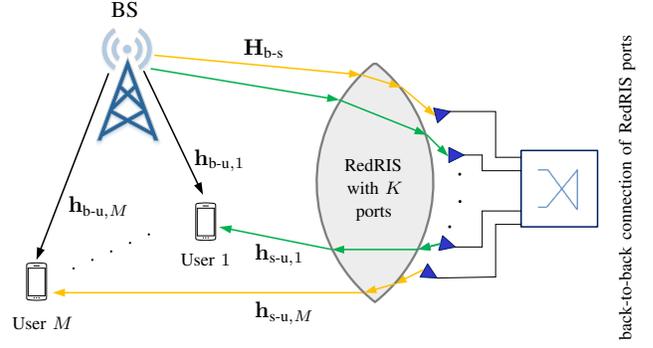

Consider a communication system consisting of one multi-antenna BS\footnote{For convenience, we focus in this section on the single-cell scenario. Extending the results to a multi-cell network will be done later on in Section \ref{sec:multiple-BS}.}, a lens-type RedRIS, and $M$ single-antenna users. The BS is equipped with $N$ ($N>M$) antenna elements. The lens-type RedRIS with $K$ ($K>M$) antenna ports is placed between the BS and the users to assist the communication. Based on CSI knowledge, back-to-back connections are established between appropriately selected combinations of unique pairs among the RedRIS ports. Fig. \ref{fig:sys-model} illustrates the system model in which there exists a direct link between the BS and each $m$-th user, represented by a channel vector $\hs_{\tn{b-u},m} \in \C^{N}$. The channel vector for the link between the RedRIS and the $m$-th user is denoted as $\hs_{\tn{s-u},m} \in \C^{K}$. Let the MIMO channel between the BS and the RedRIS be denoted as $\h?{b-s} \in \C^{K\times N}$ with Rank$(\h?{b-s})\geq M$. The incident signal at the lens-type RedRIS is redirected to the intended users owing to the refractive property of the lens and the reconfigured back-to-back connections of the antenna ports. The $K\times K$ switching matrix of the RedRIS ports is a symmetric permutation matrix denoted as $\mathbf{\Upsilon}$. A non-zero element $\upsilon_{ik}$ in $\mathbf{\Upsilon}$ means that ports $i$ and $k$ $(i \neq k)$ are back-to-back connected. We set the diagonal elements of $\mathbf{\Upsilon}$ to zero to avoid self-reflection on the same RedRIS port. Being a Fourier plane, the lens performs spatial Fourier transform on both the incident and refracted signals. The signal received by the $m$-th user, {$m \in \{1,\ldots,M\}$}, is thus expressed as follows:

\begin{eqnarray} \label{1}
    \ysn_m ~=~ \hs^\hr_{\textnormal{s-u},m}\U\mathbf{\Upsilon}\U\h?{b-s} \sum_{m'=1}^{M} \fs_{m'}\ssf_{m'}\,
    &\\ \nonumber& \!\!\!\!\!\!\!\!\!\!\!\!\!\!\!\!\!\!\!\!\!\!\!\!\!\!\!\!\!\!\!\!\!
    +~ \hs^\hr_{\textnormal{b-u},m} \sum_{m'=1}^{M} \fs_{m'}\ssf_{m'} +~ \wsn_m.
\end{eqnarray}



\noindent Here, $\U \in \C^{K \times K}$ is the $2$D DFT matrix obtained as the Kronecker product of a smaller dimension DFT matrix $\U_1 \in \C^{\sqrt{K} \times \sqrt{K}}$ with itself \cite{ghouse19932d}:

\begin{center}
    $\U = \U_1 \otimes \U_1$.
\end{center} 
Moreover, $\ssf_m \sim \CG\!\N(s_m;0,1)$ is the unknown transmitted symbol, $\wsn_m \sim \CG\!\N(w_m;0,\sigma^2?w)$ is the additive white Gaussian noise (AWGN). Besides, $\fs_{m} \in \C^N$ for $m=1,\ldots,M$ denote the active precoding vectors which are used for power allocation and beamforming purposes at the BS. Let $P$ be the total transmit power and $\f = [\fs_1,\fs_2,\ldots,\fs_M]$ be the overall precoding matrix. It follows that $\E \{ \norm {\f\ssn}^2_2 \} = P$ with $\ssn = [\ssf_1, \ssf_2, \ldots, \ssf_M]^\tr$ stacking the $M$ unknown symbols transmitted by all the users. By further defining $\ys \triangleq [\ysn_1, \ysn_2, \ldots, \ysn_M]^\tr$, $\ws \triangleq [\wsn_1, \wsn_2,\ldots, \wsn_M]^\tr$, $\h?{b-u} \triangleq \left[\hs_{\tn{b-u},1}, \hs_{\tn{b-u},2}, \ldots, \hs_{\tn{b-u},M}\right]$, and $\h?{s-u} \triangleq [\hs_{\tn{s-u},1}, \hs_{\tn{s-u},2}, \ldots, \hs_{\tn{s-u},M}]$, then the input output relationship of the RedRIS-aided multi-user MIMO system is expressed in a compact matrix-vector form:
\begin{equation} \label{2}
\ys \,=\, \h^\hr?{s-u}\U\mathbf{\Upsilon}\U\h?{b-s} \f \ssn + \h^\hr?{b-u} \f \ssn +~ \ws.
\end{equation}
By denoting the overall effective channel matrix as:
\begin{equation}\label{3}
\h \,\triangleq\, \h^\hr?{s-u}\U\mathbf{\Upsilon}\U\h?{b-s}  + \h^\hr?{b-u},
\end{equation}
the signal received by all the users can be expressed in a more succinct form as follows:
\begin{equation}\label{4}
\ys \,=\, \h \f \ssn \,+~ \ws.
\end{equation}
The goal is to minimize the received symbols' errors of all users under the sum MMSE criterion. For a given MMSE on the received symbols, the spectral efficiency of the $m$-th user is lower-bounded by \cite{heathmimo}:
\begin{equation}\label{5}
C^{\textnormal{MMSE}}_m \,=\, \log_2 \Bigg( \frac{1}{\textnormal{MMSE}_m}\Bigg), ~~m = 1,2,\ldots,M.
\end{equation}
We maximize the spectral efficiency of all users by jointly maximizing the $M$ lower bounds in \eqref{5} or equivalently by minimizing the sum MMSE criterion as follows:
\begin{subequations} \label{6}
\begin{alignat}{2}
&\!\argmin_{\alpha,\f,\mathbf{\Upsilon}} &\qquad& \E_{\ys,\ssn}\{\norm{\alpha\ys - \ssn}^2_2\},\\
&\text{subject to} &      & \E_{\ssn} \{ \norm {\f\ssn}^2_2 \} = P,\\
&                  &      & \mathbf{\Upsilon}=\mathbf{\Upsilon}^\tr,\\
& &      & \mathbf{\Upsilon}~\textrm{is a permutation matrix},\\
&                  &      & \mathbf{\Upsilon}_{kk}=0 ~\textrm{for}~k=1,2,\ldots,K.
\end{alignat}
\end{subequations}
 Here, $\alpha \in \R$ is the receive scaling factor \footnote{Without loss of generality, $\alpha$ is taken to be real valued since any phase can be included in the precoding matrix $\mathbf{F}$.} \cite{8081330,lagrange}. By recalling the expression of $\ys$ in \eqref{2} and resorting to some algebraic manipulations, it can be shown that the optimization problem in \eqref{6} is equivalent to:
\begin{subequations} \label{7}
\begin{alignat}{2}
\nonumber &\!\argmin_{\alpha,\f,\mathbf{\Upsilon}} &\qquad& \!\!\!\!\!\!\!\!\!\! \norm{\alpha\h^\hr?{s-u}\U\mathbf{\Upsilon}\U\h?{b-s}\f-(\I_M-\alpha\h^\hr?{b-u}\f)}^2_\textnormal{F}\\ 
& &   &+M\alpha^2\sigma^2_w,\\
&\text{subject to} &      & \!\!\!\!\!\!\!\!\!\! \norm {\f}^2_\textnormal{F} = P,\\
&                  &      & \!\!\!\!\!\!\!\!\!\! \mathbf{\Upsilon}=\mathbf{\Upsilon}^\tr,\\
& &      & \!\!\!\!\!\!\!\!\!\! \mathbf{\Upsilon}~\textrm{is a permutation matrix},\\
&                  &      & \!\!\!\!\!\!\!\!\!\! \mathbf{\Upsilon}_{kk}=0 ~\textrm{for}~k=1,2,\ldots,K. \label{eq7e}
\end{alignat}
\end{subequations}

\noindent The optimization problem in \eqref{7} is non-convex due to the permutation constraint on $\mathbf{\Upsilon}$ \footnote{{The constraint \ref{eq7e} is needed if an amplifier is introduced within the port connection for better gain \cite{mezghani2022nonlocal}.}}. We aim to solve it using an alternating minimization-based algorithm.

\section{Constrained Optimization of Symmetric Permutation Matrix}

\label{sec:algorithm}
Let $\mathcal{P}_K$ {be the set of {all possible symmetric permutation matrices of size $K\times K$}} and consider the related quadratic optimization problem:
\begin{subequations} \label{8}
\begin{alignat}{2}
&\!\argmin_{\x \in \R^{K \times K}} &\qquad& \norm{\A\x\B - \z}^2_\textnormal{F}, \label{8a}\\
&\text{subject to} &      & \x=\x^\tr,\\
& &      & \x ~\textrm{is a permutation matrix}, \\
&                  &      & \x_{kk}=0 ~\textrm{for}~k=1,2,\ldots,K.
\end{alignat}
\end{subequations}
in which the three matrices $\A \in \C^{M\times K}$, $\B \in \C^{K\times N}$, and $\z\in \C^{M\times N}$ are known, while $\x \in \mathcal{P}_K $ is to be optimized.
%
The underlying  optimization problem is non-convex and will be solved (sub-optimally) by alternating between  $i)$ a regularized least-squares (R-LS) module that solves the unconstrained problem and $ii)$ a projection module that enforces the constraints as shown in Fig. \ref{block1}.
\subsubsection{ R-LS Module} 
Given an extrinsic update, $\bar{\x}$, from the projection module we consider the following regularized least-squares optimization problem:
\begin{equation}
\widetilde{\x}_{\textrm{R-LS}} ~=~\argmin_{\x \in \R^{K \times K}} \qquad \norm{\A\x\B - \z}^2_\textnormal{F} +\gamma_0\|\x-\bar{\x}\|^2_\textnormal{F}, \label{RLS}
\end{equation}
wherein $\gamma_0$ is a regularization parameter to be tuned. By vectorizing the cost function in (\ref{RLS}) and  defining  $\mathbf{C} \triangleq (\B^\tr\otimes\A) \in \C^{MN \times K^2}$ it follows that:

\begin{eqnarray}
    \widetilde{\x}_{\textrm{R-LS}} ~= &\\ \nonumber& \!\!\!\!\!\!\!\!\!\!\!\!\!\!\!\!
    \tn{unvec}\left(\left( \mathbf{C}^\hr\mathbf{C}+\gamma_0\I_{K^2}\right)^{-1} \left(\mathbf{C}^\hr \tn{vec}\left(\z\right)+\gamma_0\textrm{vec}(\bar{\x})\right)\right).
\end{eqnarray}
The associated  precision is given by:
\begin{equation}
  \widetilde{\gamma}_{\textrm{R-LS}} ~=~\frac{K^2}{ \textrm{tr}\left((\mathbf{C}^\hr\mathbf{C}+\gamma_0\I_{K^2})^{-1}\right)},
\end{equation}
from which one computes the extrinsic precision, $\widetilde\gamma$, that will be fed back to the projection module:
\begin{equation}
\widetilde\gamma~ =~\widetilde{\gamma}_{\textrm{R-LS}} -\gamma_0. 
\end{equation}
The extrinsic update, $\widetilde{\x}$, that {is} provided to the projection module is updated as follows:
\begin{equation}
    \widetilde{\x} ~=~\widetilde\gamma^{-1}\left(\widetilde{\gamma}_{\textrm{R-LS}}\widetilde{\x}_{\textrm{R-LS}}-\gamma_0\bar{\x}\right).
\end{equation}

\subsubsection{Projection Module}
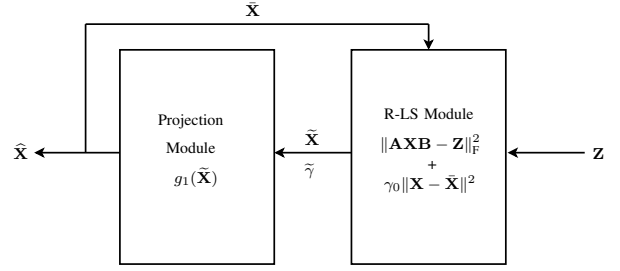
\begin{figure}
\bc
\scalebox{0.75}{
\begin{picture}(350,150)
\put(30,0){\includegraphics[scale=0.6]{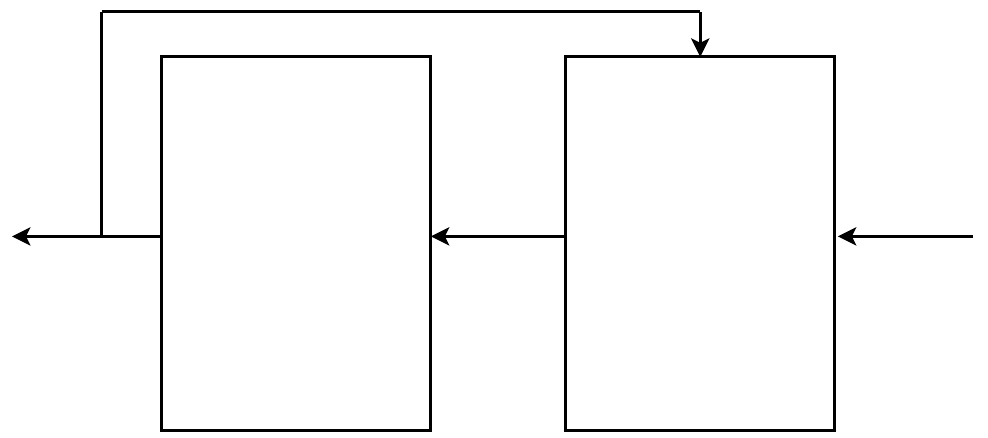}}
\put(96,72){{\footnotesize Projection}}
\put(100,58){{\footnotesize Module}}
\put(104,43){{\footnotesize  $g_1(\widetilde\x)$}}
\put(210,75){{\footnotesize R-LS Module}}
\put(208,60){{\footnotesize $\norm{\A\x\B - \z}^2_\textnormal{F} $}}
\put(232,50){{\footnotesize +}}
\put(210,40){{\footnotesize $\gamma_0\|\x-\bar{\x}\|^2$}}
\put(23,55){{\footnotesize $\widehat{\x}$}}
\put(140,127){{\footnotesize $\bar{\x}$}}
\put(170,47){{\footnotesize $\widetilde{\gamma}$}}
\put(315,55){{\footnotesize $\z$}}
\put(170,62){{\footnotesize $\widetilde{\x}$}}
\end{picture}
}
\ec
\caption{Block diagram of the proposed algorithm for symmetric permutation matrix optimization.} 
\label{block1}
\end{figure}

The projection module enforces the constraints by finding the nearest symmetric permutation matrix, $\widehat{\x}$, to the regularized least-squares solution, $\widetilde{\x}$, i.e.:


\begin{subequations} \label{11}
\begin{alignat}{2}
\widehat{\x}&~=~ &\!&\argmin_{\x\in\mathcal{P}_K} ~ \norm{\widetilde{\x} - \x}^2_\textnormal{F}~=~\argmax_{\x\in\mathcal{P}_K} ~ \textnormal{Tr}(\x\Re\{\widetilde{\x}\}), \label{11_b}\\
&\text{s.t.} &      & \x=\x^\tr,\\
&                  &      & \x_{kk}=0 ~\textrm{for}~k=1,2,\ldots,K. \label{11_c}
\end{alignat}
\end{subequations}
\noindent  
%
The optimization in \eqref{11_b} is an assignment-like problem which can be efficiently solved using the Hungarian algorithm \cite{hungarian} with cubic complexity, i.e., $\mathcal{O}(K^3)$. For simplicity, however, we develop a simpler greedy approach that is shown in \textbf{Algorithm \ref{algo:projector}} where the objective function in (\ref{11_b}) is maximized sequentially. For convenience, in the sequel we use a shorthand notation for the projector function that obtains $\widehat{\x}$ from $\widetilde{\x}$ according to \textbf{Algorithm \ref{algo:projector}}, i.e.:
\begin{equation}\label{14}
  \widehat{\x}=g_1(\widetilde{\x}).
\end{equation}
The extrinsic update $\bar{\x}$ that will be provided to the regularized least-square module is computed as follows:
\begin{equation}
  \bar{\x}  ~=~\gamma_0^{-1}\left(\big[\gamma_0+\widetilde{\gamma}\big]\widehat{\x}-\widetilde{\gamma}\widetilde{\x}\right).
\end{equation}
\noindent The process of first finding $\widetilde{\x}$ and then projecting it on the set of symmetric permutation matrices, $\mathcal{P}_k$, so as to obtain $\widehat{\x}$ is repeated until convergence. The block diagram and the algorithmic steps for the optimization framework are shown in Fig. \ref{block1} and summarized in \textbf{Algorithm \ref{algo:opt-vamp}}, respectively. 
\begin{algorithm}
\caption{Greedy algorithm for the projection module}
\label{algo:projector}
Given $\widetilde{\x} \in \C^{K \times K}$
\begin{algorithmic}[1]
\STATE $\widetilde{\x}^{(1)}= \widetilde{\x} - \textnormal{Diag} (\widetilde{\x})$
\STATE Initialize $\mathcal{S}=\{(i,j) \mid i = 1,\ldots,K;~j = 1,\ldots,K;~i \neq j\}$ and $\widehat{\x}=\mathbf{0}_K$
\WHILE{${\mathcal{S}} \neq \varnothing$}
\STATE $(\widehat{i},\widehat{j}) = \argmax\limits_{(i,j)\in \mathcal{S}}~\Re\{\widetilde{\x}^{(1)}_{ij}\}+\Re\{\widetilde{\x}^{(1)}_{ji}\}$
\STATE $\widehat{\x}_{\widehat{i}\,\,\widehat{j}}=1, \widehat{\x}_{\widehat{j}\,\,\widehat{i}}=1$
\STATE ${\mathcal{S}} \leftarrow {\mathcal{S}}\setminus \big\{(i,j) \mid i=\widehat{i} \textrm{ or } j=\widehat{j}\big\}$
\ENDWHILE
\RETURN $\widehat{\x}$
\end{algorithmic}
\end{algorithm}

\section{Optimization of RedRIS-Aided Beamforming}

\label{sec:solution}

In this section, we simultaneously optimize the ports switching matrix of the lens-type RedRIS, $\mathbf{\Upsilon}$, the BS precoding matrix, $\f$, and the receive scaling factor, $\alpha$. The proposed {solution} alternates between \textbf{Algorithm \ref{algo:opt-vamp}} to find the optimal $\mathbf{\Upsilon}$ and Lagrange optimization to find the optimal $\f$ and $\alpha$. To do so, we first split the joint optimization problem in \eqref{6} into two simpler sub-optimization tasks as follows:

\begin{subequations} \label{15}
\begin{alignat}{2}
&\!\argmin_{\mathbf\Upsilon} &\qquad& \E_{\ys,\ssn}\{\norm{\alpha\ys - \ssn}^2_2\},\\
&\text{subject to} &      & \mathbf{\Upsilon}=\mathbf{\Upsilon}^\tr,\\
& &      & \mathbf{\Upsilon}~\textrm{is a permutation matrix},\\
&                  &      & \mathbf{\Upsilon}_{kk}=0 ~\textrm{for}~k=1,2,\ldots,K.
\end{alignat}
\end{subequations}
\begin{subequations} \label{16}
\begin{alignat}{2}
&\!\argmin_{\alpha,\f} &\qquad& \E_{\ys,\ssn}\{\norm{\alpha\ys - \ssn}^2_2\},\qquad \qquad~~\\ 
&\text{subject to} &      & \E_{\ssn} \{ \norm {\f\ssn}^2_2 \} = P.             
\end{alignat}
\end{subequations}

\begin{algorithm}
\caption{Algorithm for the optimization of symmetric permutation matrices}
\label{algo:opt-vamp}
Given $\A\in\C^{M \times K}$, $\B \in \C^{K \times N}$, $\z \in \C^{M \times N}$, required convergence precision $(\epsilon)$ and maximum number of iterations $(T?{max})$
\begin{algorithmic}[1]
\STATE Initialize $\bar{\x}_0$, $\gamma_0 \geq 0$ and $t \leftarrow 1$
\REPEAT
\STATE \vspace{2pt}// R-LS Module.
\STATE $\mathbf{C} = (\B^\tr \otimes \A)$
\STATE {$\widetilde{\x}_{\textrm{R-LS},t} = \tn{unvec}\Big(\big( \mathbf{C}^\hr\mathbf{C}+\gamma_0\I_{K^2}\big)^{-1} \big(\mathbf{C}^\hr \tn{vec}\left(\z\right)$ \\ $~~~+~\gamma_0\textrm{vec}(\bar{\x}_{t-1})\big)\Big)$}
\STATE $\widetilde{\gamma}_{\textrm{R-LS}} =\frac{K^2}{ \textrm{tr}\left((\mathbf{C}^\hr\mathbf{C}+\gamma_0\I_{K^2})^{-1}\right)}$
\STATE $\widetilde\gamma = \widetilde{\gamma}_{\textrm{R-LS}} -\gamma_0$
\STATE $\widetilde{\x}_t =\widetilde\gamma^{-1}\left(\widetilde{\gamma}_{\textrm{R-LS}}\widetilde{\x}_{\textrm{R-LS},t}-\gamma_0\bar{\x}_{t-1}\right)$
\STATE // Projection Module.
\STATE Compute projected matrix $\widehat{\x}_t=g_1(\widetilde{\x}_t)$ using \textbf{Algorithm \ref{algo:projector}}
\STATE $\bar{\x}_t  =\gamma_0^{-1}\left(\big[\gamma_0+\widetilde{\gamma}\big]\widehat{\x}_t-\widetilde{\gamma}\widetilde{\x}_t\right)$
\STATE $t\leftarrow t+1$
\vspace{2pt}
\UNTIL $\norm{\widehat{\x}_t-\widehat{\x}_{t-1}}?F^2\leq \epsilon\norm{\widehat{\x}_{t-1}}?F^2$ or $t>T?{max}$
\RETURN $\widehat{\x}_t$
\end{algorithmic}
\end{algorithm}

\subsection{Finding the optimal switching matrix of the RedRIS ports}

We use the R-LS and projection modules discussed in Section \hyperref[sec:algorithm]{III} to solve the sub-optimization problem in \eqref{15} which is explicitly re-stated as:
\begin{subequations} \label{17}
\begin{alignat}{2}
&\!\argmin_{\mathbf\Upsilon} &\qquad& \norm{\alpha\h^\hr?{s-u}\U\mathbf{\Upsilon}\U\h?{b-s}\f-(\I_M-\alpha\h^\hr?{b-u}\f)}^2_\textnormal{F},\\
&\text{subject to}                 &      & \mathbf{\Upsilon}=\mathbf{\Upsilon}^\tr,\\
& &      & \mathbf{\Upsilon}~\textrm{is a permutation matrix},\\
&                  &      & \mathbf{\Upsilon}_{kk}=0 ~\textrm{for}~k=1,2,\ldots,K.
\end{alignat}
\end{subequations}

\noindent We obtain the solution by substituting $\A=\alpha\h^\hr?{s-u}\U$, $\B = \U\h?{b-s}\f$, and $\z = \I_M-\alpha\h^\hr?{b-u}\f$ in \eqref{8a} and solve for $\x = \mathbf\Upsilon$ using \textbf{Algorithm \ref{algo:opt-vamp}} in Section \hyperref[sec:algorithm]{III}.
\subsection{Finding the optimal BS precoding matrix and receive scaling factor}

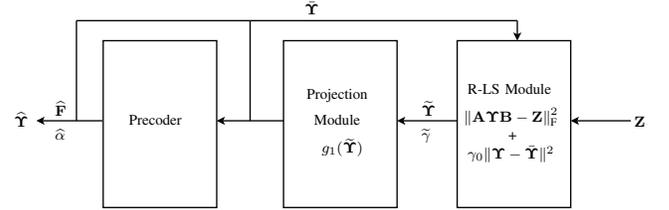
\begin{figure}
\bc
\scalebox{0.7}{
\begin{picture}(370,140)
\put(22,12){\includegraphics[scale=0.5]{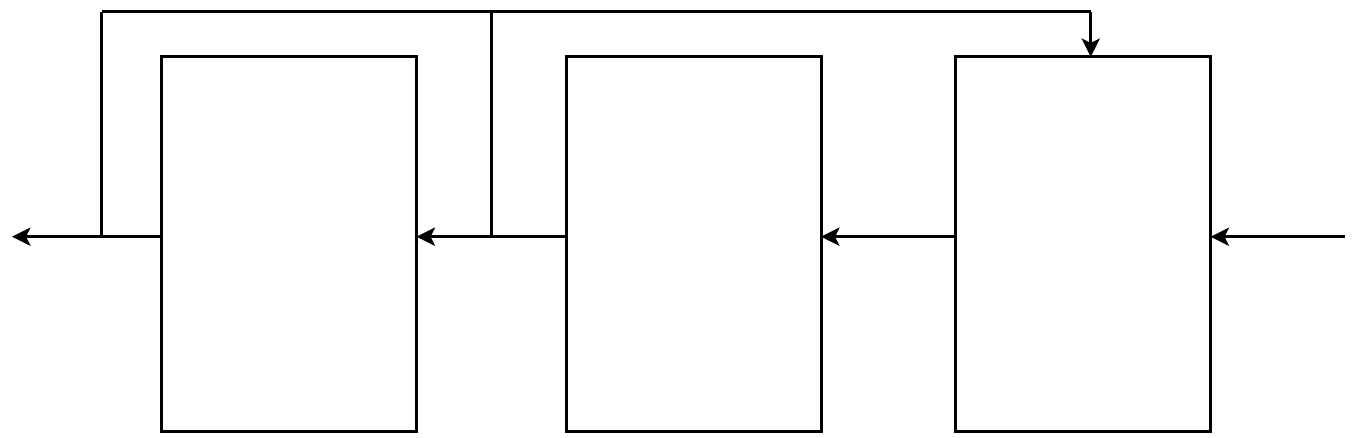}}
\put(171,72){{\footnotesize Projection}}
\put(175,58){{\footnotesize Module}}
\put(179,43){{\footnotesize  $g_1(\widetilde{\mathbf{\Upsilon}})$}}
\put(258,75){{\footnotesize R-LS Module}}
\put(256,60){{\footnotesize $\norm{\A\mathbf{\Upsilon}\B - \z}^2_\textnormal{F} $}}
\put(280,50){{\footnotesize +}}
\put(258,40){{\footnotesize $\gamma_0\|\mathbf{\Upsilon}-\bar{\mathbf{\Upsilon}}\|^2$}}
\put(13,57){{\footnotesize $\widehat{\mathbf{\Upsilon}}$}}
\put(170,118){{\footnotesize $\bar{\mathbf{\Upsilon}}$}}
\put(233,50){{\footnotesize $\widetilde{\gamma}$}}
\put(348,57){{\footnotesize $\z$}}
\put(233,63){{\footnotesize $\widetilde{\mathbf{\Upsilon}}$}}
\put(75,58){{\footnotesize Precoder}}
\put(35,63){{\footnotesize $\widehat{\f}$}}
\put(35,50){{\footnotesize $\widehat{\alpha}$}}
\end{picture}
}
\ec
\caption{Block diagram of the proposed algorithm.}
\label{block diagram}
\end{figure}

The sub-optimization task in \eqref{16} is a constrained MMSE transmit precoding design problem for traditional MIMO systems which can be solved by jointly optimizing $\f$ and $\alpha$ using Lagrange optimization. The optimal $\alpha$ and $\f$ are obtained in closed-form expressions by following the same approach developed in \cite{vampmain}: 
\begin{subequations} \label{18}
\begin{alignat}{2}
&\widehat{\alpha} & \,=\, & g_2\left(\h\right)\triangleq\mathsmaller{\sqrt{\frac{1}{P}}}\sqrt{\tn{Tr}\left(\left[\h^\hr\h+\mathsmaller{\frac{M\sigma?w^2}{P}}\I_N\right]^{-2}\h^\hr\h\right)}, \\ 
&\widehat{\f} & \,=\, & g_3\left(\h\right)\triangleq\frac{1}{\widehat{\alpha}}\left(\h^\hr\h+\mathsmaller{\frac{M\sigma?w^2}{P}}\I_N\right)^{-1}\h^\hr.             
\end{alignat}
\end{subequations}
A detailed description of the solution can be found in \cite{lagrange}. At this stage, the MSE at iteration $t$ is given by:
\begin{equation}\label{20}
    E_t ~\triangleq~ \norm{\widehat{\alpha}_t(\h^\hr?{s-u}\U\widehat{\mathbf\Upsilon}_t\U\h?{b-s}+\h^\hr?{b-u})\widehat{\f}_t-\I_M}^2_\textnormal{F} + M\widehat{\alpha}_t^2\sigma^2?w.
\end{equation}
The algorithm stops iterating when $\abs{E_{t}-E_{t-1}} < \epsilon E_{t-1}$, where $\epsilon \in \R_+$ is some required convergence precision. The overall block diagram and the algorithmic steps are shown in Fig. \ref{block diagram} and \textbf{Algorithm \ref{algo:overall}}. The R-LS module of the algorithm can be efficiently implemented by exploiting the Kronecker matrix structure and singular value decomposition (SVD) \cite{vampmain}.

\begin{algorithm}
\caption{Alternating optimization of $\mathbf{\f}$, $\alpha$, and $\mathbf{\Upsilon}$}
\label{algo:overall}
Given $\h?{s-u}$, $\h?{b-u}$, $\h?{b-s}$, $K \times K$ DFT matrix $\mathbf{(U)}$, required convergence precision $(\epsilon)$, and maximum number of iterations $(T?{max})$
\begin{algorithmic}[1]
\STATE Initialize $\widehat{{\mathbf\Upsilon}}?0$, $\bar{\mathbf\Upsilon}_0$, $\gamma_0 \geq 0$ and $t\leftarrow 1$
\STATE $\widehat{\h}_0=\h?{s-u}^\hr\U\widehat{{\mathbf\Upsilon}}?0\U\h?{b-s} +\h?{b-u}^\hr$
\STATE $\widehat{\alpha}?0=g_2\left(\widehat{\h}_0\right)$
\STATE $\widehat{\f}?0=g_3\left(\widehat{\h}_0\right)$ \vspace{2pt}
\REPEAT
\STATE // R-LS Module.
\STATE $\A= \widehat{\alpha}_{t-1}\h?{s-u}^\hr\U$ 
\STATE $\B=\U\h?{b-s}\widehat{\f}_{t-1}$.
\STATE $\z=\I_M-\widehat{\alpha}_{t-1}\h?{b-u}^\hr\widehat{\f}_{t-1}$
\STATE $\mathbf{C} = (\B^\tr \otimes \A)$
\STATE {$\widetilde{\mathbf\Upsilon}_{\textrm{R-LS},t} = \tn{unvec}\Big(\big( \mathbf{C}^\hr\mathbf{C}+\gamma_0\I_{K^2}\big)^{-1} \big(\mathbf{C}^\hr \tn{vec}\left(\z\right)$ \\ $~~~+~\gamma_0\textrm{vec}(\bar{\mathbf\Upsilon}_{t-1})\big)\Big)$}
\STATE $\widetilde{\gamma}_{\textrm{R-LS}} =\frac{K^2}{ \textrm{tr}\left((\mathbf{C}^\hr\mathbf{C}+\gamma_0\I_{K^2})^{-1}\right)}$
\STATE $\widetilde\gamma = \widetilde{\gamma}_{\textrm{R-LS}} -\gamma_0$
\STATE $\widetilde{\mathbf\Upsilon}_t =\widetilde\gamma^{-1}\left(\widetilde{\gamma}_{\textrm{R-LS}}\widetilde{\mathbf\Upsilon}_{\textrm{R-LS},t}-\gamma_0\bar{\mathbf\Upsilon}_{t-1}\right)$
\STATE // Projection Module.
\STATE Compute the projected matrix $\widehat{\mathbf\Upsilon}_t=g_1(\widetilde{\mathbf\Upsilon}_t)$ using \textbf{Algorithm \ref{algo:projector}}
\STATE $\bar{\mathbf\Upsilon}_t  =\gamma_0^{-1}\left(\big[\gamma_0+\widetilde{\gamma}\big]\widehat{\mathbf\Upsilon}_t-\widetilde{\gamma}\widetilde{\mathbf\Upsilon}_t\right)$
\STATE // Find the optimal $\alpha$ and $\f$.
\STATE $\h_t= \h?{s-u}^\hr\U\widehat{{\mathbf\Upsilon}}_t\U\h?{b-s} +\h?{b-u}^\hr$
\STATE $\widehat{\alpha}_t=g_2\big(\widehat{\h}_t\big)$
\STATE $\widehat{\f}_t=g_3\big(\widehat{\h}_t\big)$
\STATE $t \leftarrow t+1$
\UNTIL $|E_t-E_{t-1}|<\epsilon E_{t-1}$ or $t>T?{max}$
\RETURN $\widehat{\f}_t, \ \widehat{\alpha}_t, \ \widehat{{\mathbf\Upsilon}}_t$.
\end{algorithmic}
\end{algorithm}


\section{Reducing the number of Port Switchings}

\label{sec:reduction}

The dimension of the switching matrix, $\mathbf{\Upsilon}$, increases with the number of antenna ports in the lens-type RedRIS, resulting in an increased signaling overhead from the BS to the RedRIS. In this section, we propose two methods for reducing the number of connected ports down to $N_p$ $(N_p\ll K)$  ports while incurring a negligible degradation in overall performance. In a large-sized RedRIS, not all the beams in the incident wave are required to successfully reconstruct the signal at the receiver end. {This is because} the main beam and a few adjacent beams are usually sufficient to decode the transmitted information. The adjacent beams are required because of the leakage effect. This can be exploited to dynamically select the most significant ports so as to steer the RedRIS towards the dominant beams only. In this way, back-to-back connections are established between the selected antenna ports only, while leaving the other ports disconnected. Reducing the number of the RedRIS ports connections significantly reduces the incurred control overhead and the overall complexity of the system. We propose two reduction approaches as outlined in the sequel.

\subsection{Universal reduction}
In presence of a RedRIS with a large number of antenna ports {$(K)$} a large-size $K\times K$ switching matrix, $\mathbf\Upsilon$, establishes $K/2$ back-to-back port-connections. To obtain a reduced-size switching matrix, a small number of effective ports are selected to steer the lens-type RedRIS towards the dominant beams only. We use the MMSE criterion to determine the so-called effective RedRIS ports as follows. We define the support $\mathcal{E}(\mathbf{A})$ of any symmetric matrix $\mathbf{A}$ (with zero diagonal elements) as the set of positions pertaining to non-zero entries in the lower triangular part of  $\mathbf{A}$, i.e.:
\begin{equation}
    \mathcal{E}(\mathbf{A})=\big\{(i,j)|\mathbf{A}_{ij}\neq0, j<i\big\}
\end{equation}
Let $\widehat{\mathbf{\Upsilon}}$ be the  full optimal switching matrix returned by \textbf{Algorithm \ref{algo:overall}}
whose support has cardinality $\big|\mathcal{E}(\widehat{\mathbf{\Upsilon}})\big|=K$. For each $(i,j)\in\mathcal{E}(\widehat{\mathbf{\Upsilon}})$, {we first set the associated symmetric off-diagonal entries in $\widehat{\mathbf{\Upsilon}}$ to 0, i.e.,   $(\widehat{\mathbf{\Upsilon}}_{ij},\widehat{\mathbf{\Upsilon}}_{ji}) = (0,0)$, recompute $\alpha$ and $\mathbf{F}$ according to (\ref{18}) based on this new switching matrix, and then compute the corresponding MSE.}
The smallest among the $K$ computed MSEs corresponds to some position $(\widehat{i},\widehat{j})\in\mathcal{E}(\widehat{\mathbf{\Upsilon}})$. A reduced switching matrix, $\widehat{\mathbf{\Upsilon}}_1$, is obtained as follows:
\begin{equation}\label{reduction_1}
\widehat{\mathbf{\Upsilon}}_1 = \widehat{\mathbf{\Upsilon}}- (\mathbf{e}_{\,\widehat{i}}\,\mathbf{e}_{\,\widehat{j}}^{\top}~+~\mathbf{e}_{\,\widehat{j}}\,\mathbf{e}_{\,\widehat{i}}^{\top}),
\end{equation}
where $\mathbf{e}_k$ is the $k$-th canonical basis vector in $\mathbb{R}^K$, i.e., $\mathbf{e}_k$ has a single nonzero component which is equal to one at the $k$-th position. In plain English, the operation  in (\ref{reduction_1}) sets the symmetric entries in $\widehat{\mathbf{\Upsilon}}$  at positions $(\widehat{i},\widehat{j})$ and $(\widehat{j},\widehat{i})$ to zero. Since these positions correspond to the smallest incurred MSE, removing the associated back-to-back connection (by using $\widehat{\mathbf{\Upsilon}}_1$ instead of $\widehat{\mathbf{\Upsilon}}$) has the smallest impact on performance. The above-described reduction procedure is applied again to  $\widehat{\mathbf{\Upsilon}}_1$ so as to remove the next least significant back-to-back connection and obtain another reduced switching matrix $\widehat{\mathbf{\Upsilon}}_2$. The process is repeated until only the most significant $N_p\ll K$ effective ports are maintained {in which case the number of paths in the BS-RIS channel, $Q?{RIS}$, is $N_p/2$.} The algorithmic steps of the universal reduction method is shown in \textbf{Algorithm \ref{algo:reduction}}.

\begin{algorithm}
\caption{Universal reduction of the required back-to-back connections}
\label{algo:reduction}
Given full optimal switching matrix $\mathbf{\widehat{\Upsilon}} \in\mathbb{R}^{K \times K}$, $\h?{s-u}$, $\h?{b-u}$, $\h?{b-s}$, $K \times K$ DFT matrix $\mathbf{(U)}$, number of effective ports $N_p$
\begin{algorithmic}[1]
\STATE Initialize $t\leftarrow 0$, ~~$\mathbf{\widehat{\Upsilon}}_0 = \mathbf{\widehat{\Upsilon}}$
\REPEAT
\STATE
\hspace{-0.4cm}
{ \small{
$(\widehat{i},\widehat{j}) = \argmin\limits_{(i,j)\in\mathcal{E}(\widehat{\mathbf{\Upsilon}}_t)} \norm{{\widehat{\alpha}_{t}}(\h^\hr?{s-u}\U\widehat{\mathbf\Upsilon}_t\U\h?{b-s}+\h^\hr?{b-u}){\widehat{\f}_t}-\I_M}^2_\textnormal{F},$
\begin{align*}
    \tn{s.t.} \quad &[\widehat{\mathbf{\Upsilon}}_t]_{ij} \leftarrow 0,  ~[\widehat{\mathbf{\Upsilon}}_t]_{ji} \leftarrow 0,
    \\ & \widehat{\alpha}_{t} = g_2(\h^\hr?{s-u}\U\widehat{\mathbf\Upsilon}_t\U\h?{b-s}  + \h^\hr?{b-u}), 
    \\ & \widehat{\f}_t = g_3(\h^\hr?{s-u}\U\widehat{\mathbf\Upsilon}_t\U\h?{b-s}  + \h^\hr?{b-u})
\end{align*}}
}

\vspace{-0.4cm}
\STATE $\widehat{\mathbf{\Upsilon}}_t = \widehat{\mathbf{\Upsilon}}_{t-1}- (\mathbf{e}_{\,\widehat{i}}\,\mathbf{e}_{\,\widehat{j}}^{\top}~+~\mathbf{e}_{\,\widehat{j}}\,\mathbf{e}_{\,\widehat{i}}^{\top})$\vspace{0.2cm}
\STATE $t\leftarrow t+1$
\UNTIL{$\textrm{Rank}(\mathbf{\Upsilon}?t)= N_p$}
\RETURN $\widehat{{\mathbf\Upsilon}}_t$.
\end{algorithmic}
\end{algorithm}

\subsection{Two-port solution for the single-user case}
For the single-user case $(M=1)$, an efficient approach --- referred to as the ``two-port solution'' --- that uses only $N_p=2$ effective ports at the RedRIS is proposed. The two-port solution is obtained in a closed-form by determining the two antenna elements that capture most of the energy in the cascaded BS-RedRIS-user channel. By letting $\h^{(1)}=\mathbf{U}\h_{\tn{b-s}} \in \C^{K\times N}$ and $\hs^{(2)}=\hs_{\tn{s-u}}\mathbf{U} \in \C^{1\times K}$, then the antenna-wise captured energies in the BS-RedRIS channels $({\mathbf{e}}?{b-s})$, and the user-RedRIS channels $({\mathbf{e}}?{s-u})$ are defined respectively as:
\begin{equation}
\label{22}
    {e}?{{b-s},\textit{i}} = \sum_{n =1}^N\big|\h^{(1)}_{in}\big|^2,~ i=1,\ldots,K,
\end{equation}
\begin{equation}
\label{23}
    {e}?{{s-u},\textit{j}} = \big|\hs^{(2)}_{j}\big|^2, ~~ j=1,\ldots,K.
\end{equation}
The objective function for finding the optimum port index $(\widehat{i},\widehat{j})$ for the two-port solution is expressed as follows:
\begin{subequations}
\begin{equation}
    \widehat{i} = \argmax_{1\leq i\leq K} ~~ {e}?{{b-s},\textit{i}}.
\end{equation}
\begin{equation}
     \widehat{j} = \argmax_{1\leq j\leq K} ~~ {e}?{{s-u},\textit{j}}.
\end{equation}
\end{subequations}

\section{Joint Optimization for the Multi-Cell Scenario}

\begin{table*}[t]
\centering
\caption{Simulation parameters with the notations and values.}
\label{table:sim-parameter}
\begin{tabular}{|p{0.32\textwidth}|P{0.15\textwidth}|p{0.25\textwidth}|P{0.15\textwidth}|}
\hline 
\vspace{-10pt} \center{\textbf{Parameter}}  &  \vspace{-5pt}\textbf{Notation, Value}  & \vspace{-10pt} \center{\textbf{Parameter}}  &  \vspace{-5pt}\textbf{Notation, Value}\\
 \hline
\hline 
 Multi-paths in the BS-RIS channel & $Q?{RIS}=10$ & 
 Channel path gain & $c_q\sim\CG\!\N(0,1)$ \\
\hline
 Multi-paths in the BS-user channel & $Q?{b-u}=2$ &
 Distance between BS and RIS & $d?{RIS}\in [100,500]$ m\\
\hline 
 Multi-paths in the RIS-user channel & $Q?{s-u}=2$&
 Distance between RIS and users & $d' \in [10,50]$ m \\
\hline 
 Path-loss exponent of the RIS-user channel & $\eta=2.5$&
 Noise variance & $\sigma?w^2=-100$ dBm \\
\hline 
 Path-loss exponent of the RIS-BS channel & $\eta=2.5$&
 Reference distance & $d?0=1$ m \\
\hline 
 Path-loss exponent of the BS-user channel & $\eta=3.7$&
 Path-loss at reference distance & $C?0=-30$ dB \\
\hline
\end{tabular}
\end{table*}

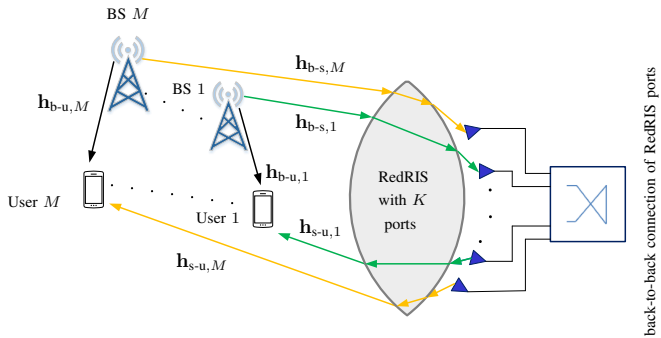
\begin{figure}
\bc
\scalebox{0.75}{
\begin{picture}(270,170)
\put(0,0){\includegraphics[scale=0.5]{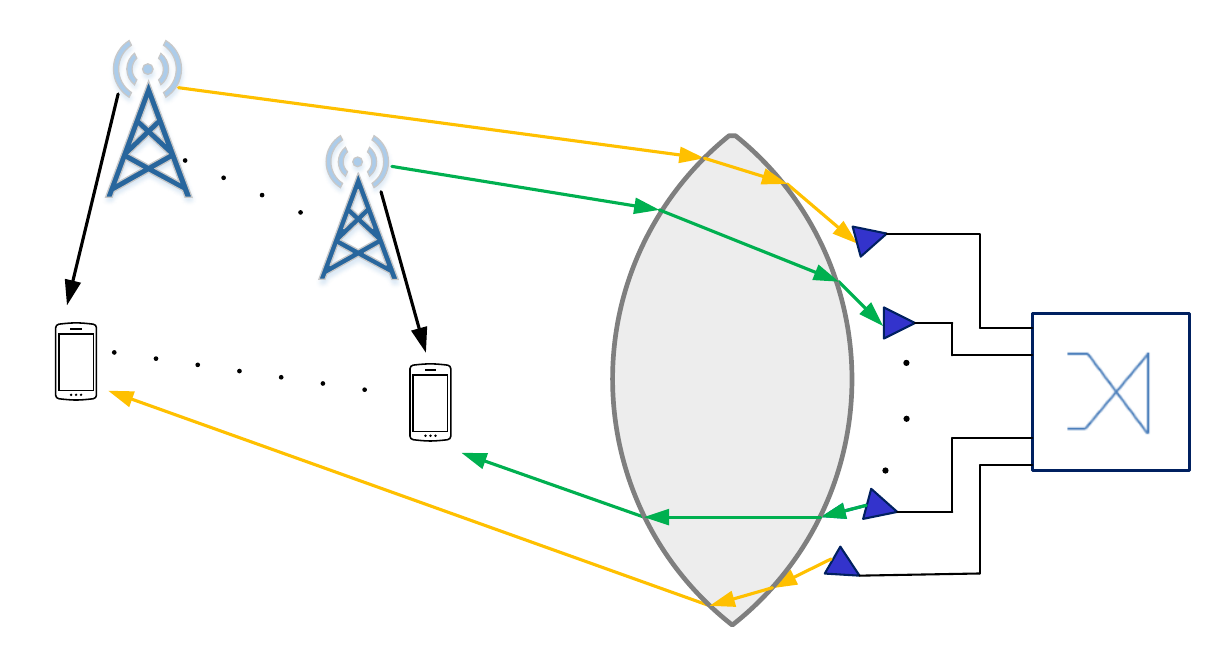}}
\put(25,160){\footnotesize{BS $M$}}
\put(58,122){\footnotesize{BS $1$}}
\put(-25,65){\footnotesize{User $M$}}
\put(70,57){\footnotesize{User $1$}}
\put(162,78){\footnotesize{RedRIS}}
\put(162,66){\footnotesize{with $K$}}
\put(165,54){\footnotesize{ports}}
\put(296,0){\rotatebox{90}{\footnotesize{back-to-back connection of RedRIS ports}}}
\put(120,105){$\hs_{\tn{b-s},1}$}
\put(120,135){$\hs_{\tn{b-s},M}$}
\put(122,52){$\hs_{\tn{s-u},1}$}
\put(60,35){$\hs_{\tn{s-u},M}$}
\put(105,80){$\hs_{\tn{b-u},1}$}
\put(-10,115){$\hs_{\tn{b-u},M}$}
\end{picture}
}
\caption{RedRIS-assisted communication {for a multi-cell} system.}
\label{fig:sys-model_extended}
\ec
\end{figure}

\label{sec:multiple-BS}
In this section, we extend the system to an $M$-cell network where each cell is equipped with a single-antenna BS and each BS is serving one user at a time \cite{jiang2022interference}. A RedRIS with $K$ ($K>M$) antenna ports is simultaneously assisting all the BSs to serve their intended users. Similar to the single-cell scenario, back-to-back connections are established between the RedRIS antenna ports to facilitate the communication. For simplicity, representation of the channel vectors is the same as in the single-cell setup (see Section \hyperref[sec:systemmodel]{II}) and Fig. \ref{fig:sys-model_extended} illustrates the system model for the multi-cell scenario. The input-output relationship for the RedRIS-aided multi-cell system is expressed as:
\begin{equation} \label{25}
\ys ~=~ \h^\hr?{s-u}\U\mathbf{\Upsilon}\U\h?{b-s} \xxn ~+~ \h^\hr?{b-u} \xxn ~+~ \ws,
\end{equation}
Here, $\h?{s-u} \in \C^{K \times M}$, $\h?{b-s} \in \C^{K \times M}$ and $\h?{b-u} \in \C^{M \times M}$ are the related channel matrices and $\xxn \triangleq \sqrt{P}\ssn$ where $P$ is the transmit power at each BS.

\noindent In \eqref{25}, no BS precoding matrix $(\f)$ is required as opposed to the single-cell case. This is because there is no joint processing between the BSs and every BS is serving its own user only. We formulate the objective function for the multi-cell scenario under the sum MMSE criterion as follows:
\begin{subequations} \label{26}
\begin{alignat}{2}
&\!\argmin_{\bm{\alpha}
,\mathbf{\Upsilon}} &\qquad& \E_{\ys,\ssn}\{\norm{\textrm{Diag}(\bm{\alpha})
\ys - \ssn}^2_2\},\\
&\text{subject to} &      & \mathbf{\Upsilon}=\mathbf{\Upsilon}^\tr,\\
& &      & \mathbf{\Upsilon}~\textrm{is a permutation matrix},\\
&                  &      & \mathbf{\Upsilon}_{kk}=0 ~\textrm{for}~k=1,2,\ldots,K.
\end{alignat}
\end{subequations}
Here, $\bm{\alpha} \in \C^{M \times 1}$ is a vector that contains the receive scaling factors of all users. By recalling the expression of $\ys$ in \eqref{25} and resorting to some algebraic manipulations, the objective function in \eqref{26} is re-written as: 
\begin{subequations} \label{27}
\begin{alignat}{2}
\nonumber &\!\!\!\!\!\!\!\!\!\!\!\!\!\!\!\!\!\!\!\!\!\!\!\!\!\!\!\!\!\!\!\! \argmin_{\bm{\alpha},\mathbf{\Upsilon}} &\qquad& \!\!\!\!\!\!\!\!\!\!\!\norm{\textrm{Diag}(\bm{\alpha})[\h^\hr?{s-u}\U\mathbf{\Upsilon}\U\h?{b-s} + \h^\hr?{b-u}]\sqrt{P}- \I_M}_\textnormal{F}^2 \\+ \sigma^2_w\|\bm{\alpha}\|^2,\\
&\!\!\!\!\!\!\!\!\!\!\!\!\!\!\!\!\!\!\!\!\!\!\!\!\!\!\text{subject to} &      & \!\!\!\!\!\!\!\!\!\!\! \mathbf{\Upsilon}=\mathbf{\Upsilon}^\tr,\\
& &      & \!\!\!\!\!\!\!\!\!\!\! \mathbf{\Upsilon}~\textrm{is a permutation matrix},\\
&                  &      &\!\!\!\!\!\!\!\!\!\!\! \mathbf{\Upsilon}_{kk}=0 ~\textrm{for}~k=1,2,\ldots,K.
\end{alignat}
\end{subequations}

The optimization problem in \eqref{27} is non-convex and will also be solved by means of \textbf{Algorithm \ref{algo:overall}} developed in Section \hyperref[sec:algorithm]{III}. For a fixed $\bm{\alpha}$, we replace $\A=\textrm{Diag}(\bm{\alpha})\h^\hr?{s-u}\U$, $\B = \U\h?{b-s}\sqrt{P}$, and $\z = \I_M-\textrm{Diag}(\bm{\alpha})\h^\hr?{b-u}\sqrt{P}$ in \eqref{8a} and solve for $\x = \mathbf\Upsilon$ by using \textbf{Algorithm \ref{algo:overall}}. {To find the optimal vector of receive scaling factors, $\bm{\alpha}$, we minimize the individual users' MSEs under the MMSE criterion as:
\begin{equation} \label{28}
\min_{\alpha_m\in\mathbb{C}} \E \big\{\abs{\alpha_m[(\hs^\hr?{s-u,m}\U\mathbf{\Upsilon}\U\h?{b-s}+\hs^\hr?{b-u,m})\sqrt{P}\ssn +\wsn_m]-\ssf_m}^2\big\}.
\end{equation}
After taking expectation with respect to $\ssn$ and $\wsn_m$, \eqref{28} reduces simply to:
\begin{equation} \label{29}
\min_{\alpha_m\in\mathbb{C}} \norm{\alpha_m \mathbf{v} - \mathbf{e}_m}^2_2+\sigma^2_w |\alpha_m|^2,
\end{equation}
wherein $\mathbf{v} \triangleq \sqrt{P}(\hs_{\textrm{s-u},m}^\hr\U\mathbf{\Upsilon}\U \h?{b-s}+\hs_{\textrm{b-u},m}^\hr)^{\mathrm{T}} \in \C^{M}$ and $\mathbf{e}_m$ is the $m$-th canonical basis vector in $\mathbb{R}^M$, i.e., $\mathbf{e}_m$ has a single nonzero component which is equal to one at the $m$-th position.
Then, by setting the derivative of the cost function in \eqref{29}  with respect to $\alpha_m$ to zero and solving for $\alpha_m$, we obtain the optimum receive scaling factor for each $m$-th user as:
\begin{equation}
\label{30}
    \alpha^{\textrm{opt}}_m = \frac{\mathbf{v}^\hr\mathbf{e}_m}{\|\mathbf{v}\|^2+\sigma^2_w} .
\end{equation}}
\noindent{The number of back-to-back connections of the RedRIS ports can be reduced by applying the approach developed in Section \hyperref[sec:reduction]{V} (i.e., \textbf{Algorithm \ref{algo:reduction}}) for further overhead savings.}
\section{Numerical results and performance analysis}
\label{sec:numeric-res}
\subsection{Simulation model and parameters}
\begin{figure*}
\centering
\begin{subfigure}{0.5\textwidth}
\includegraphics[width = \linewidth]{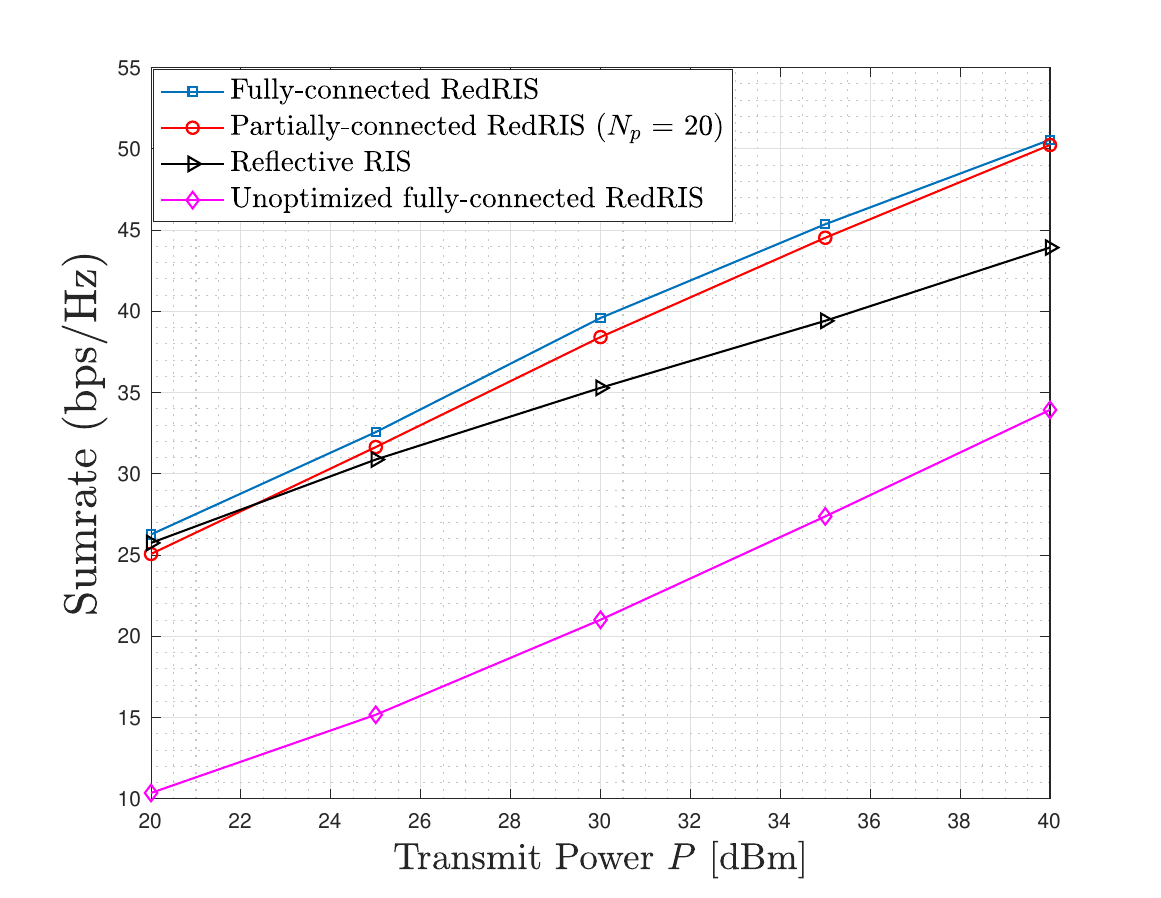}
\caption{\footnotesize $M=4, \ N=32$ and $K=256$.}
\label{figg 5a}
\end{subfigure}%
\hfill
\begin{subfigure}{.5\textwidth}
\includegraphics[width = \linewidth]{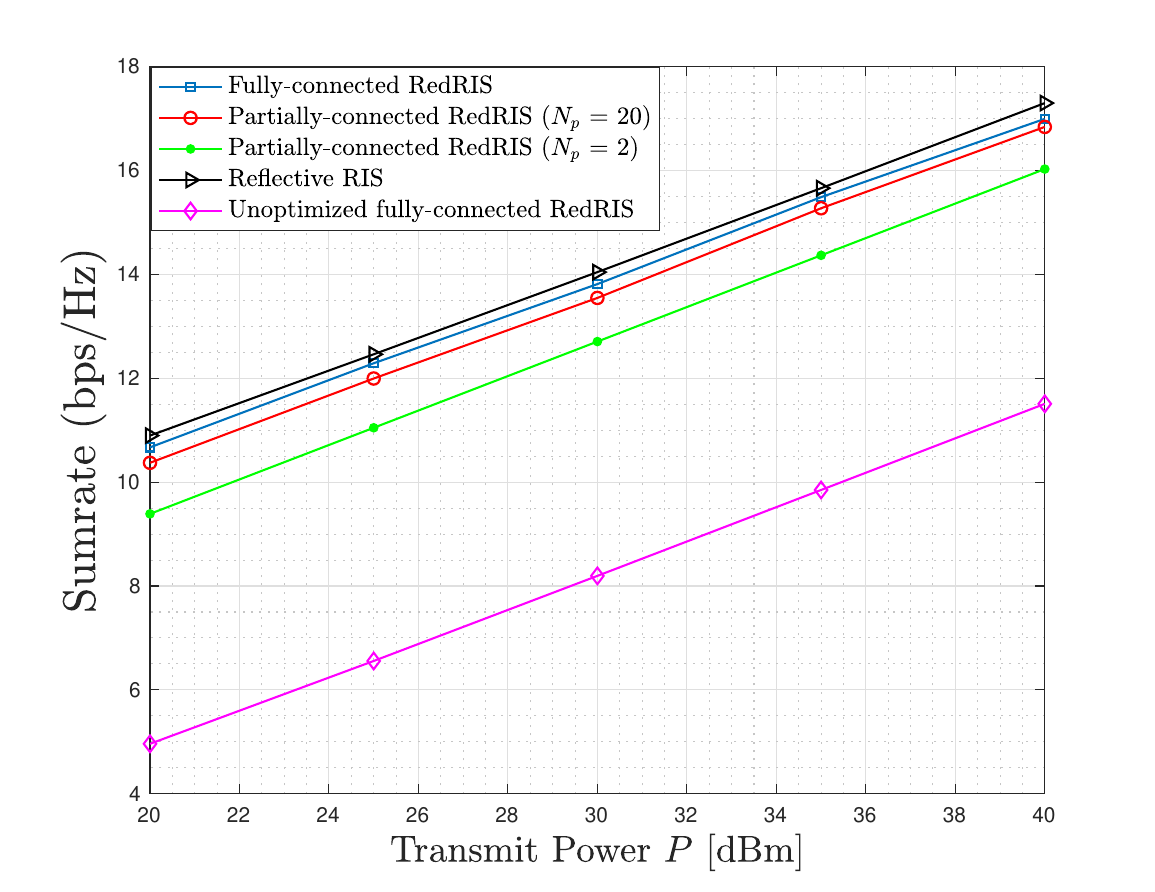}
\caption{\footnotesize $M=1, \ N=32$ and $K=256$.}
\label{figg 5b}
\end{subfigure}
\hfill
\begin{subfigure}{.5\textwidth}
\includegraphics[width = \linewidth]{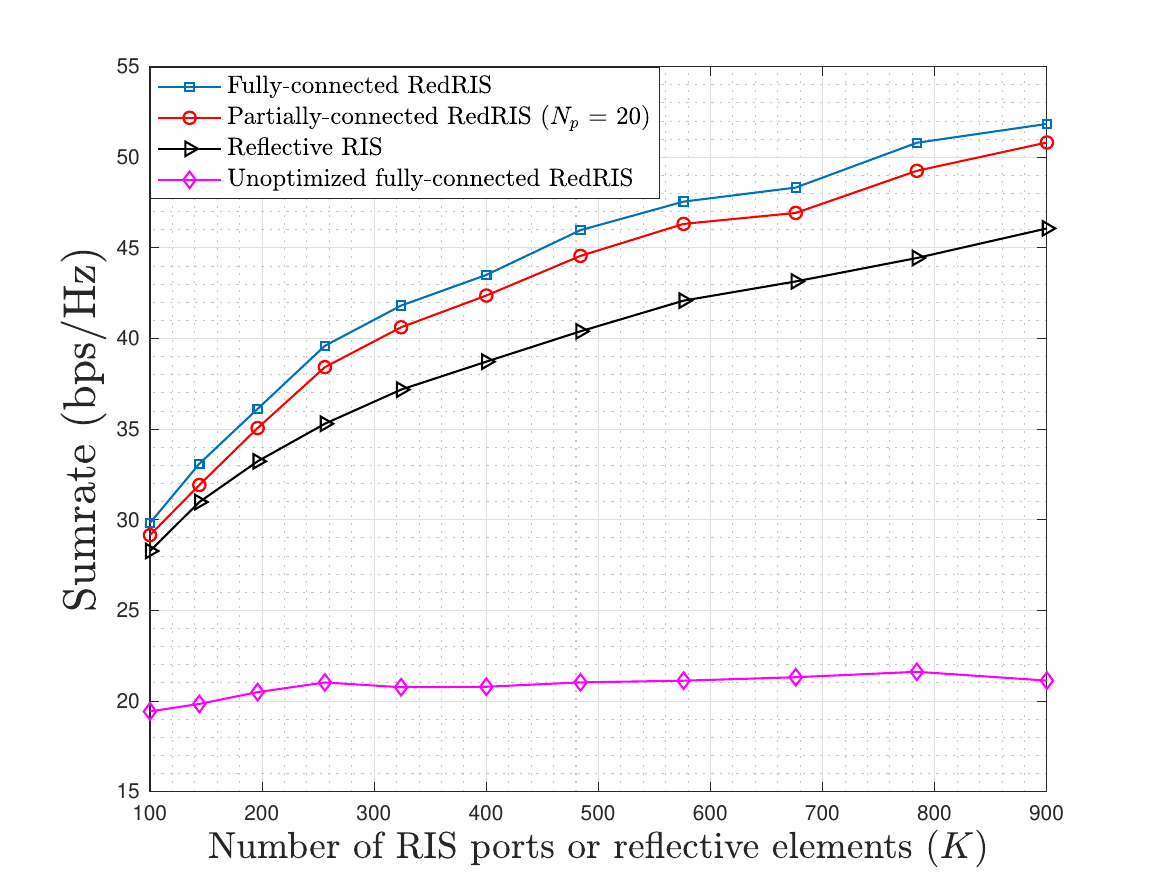}
\caption{\footnotesize $M=4, \ N=32$ and $P=30$ dBm.}
\label{figg 5c}
\end{subfigure}%
\hfill
\begin{subfigure}{.5\textwidth}
\includegraphics[width = \linewidth]{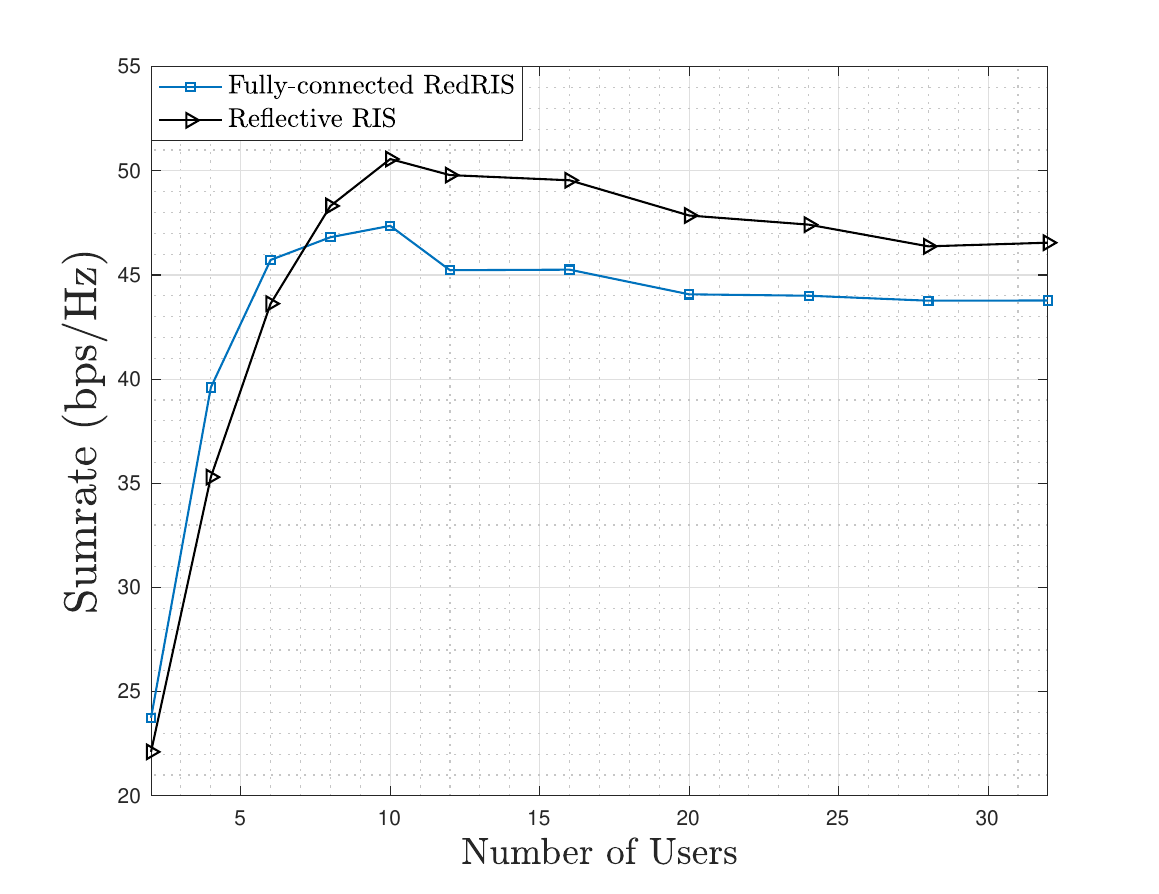}
\caption{\footnotesize $N=32, \ K=256$ and $P=30$ dBm.}
\label{figg 5d}
\end{subfigure}
\caption{(a) sum-rate versus transmit power (multi-user case), (b) sum-rate versus transmit power (single-user case), (c) sum-rate versus the number of RIS ports or reflection elements (multi-user case), and (d) sum-rate versus number of users. LOS component is present in the BS-RIS channels only.}
\end{figure*}

In this section, we present the performance of the proposed RedRIS-aided communication schemes by means of Monte-Carlo simulations and using the \textit{sum-rate} as a performance metric. Given the MMSE of the received symbol pertaining to each $m$-th user, the sum-rate of the overall system is expressed as: 
\begin{equation}
\label{31}
\widehat{C}=\ds\sum_{m=1}^M\log_2\left(\frac{1}{\textnormal{{MMSE}}_m}\right).
\end{equation}
In the single-cell scenario, we assume that the RedRIS is situated at a fixed distance $d_{\textrm{RIS}}=500$ m from the BS while in the multi-cell setup the BSs are located at a uniform radial distance of $100$ m to $500$ m from the RedRIS. In both scenarios, the users are uniformly spread at a distance ranging from $10$ m to $50$ m from the RedRIS. For simplicity purpose, we consider the users are located at $d=500$ m from their respective BSs. We assume there are $Q_{\textrm{RIS}}=10$ multi-path components in the BS-RedRIS channel and $Q_{\textrm{b-u}}=2$ multi-path components in each BS-user channel. We also assume that the $N$ antenna elements at the BS and the $K$ antenna elements at the RedRIS (connected through their respective ports) are arranged in square uniform planar arrays.
We use the parametric or path-based channel model to generate the channel matrices \cite{heathmimo}. We consider a carrier frequency of $3$ GHz with $30$ MHz bandwidth. We refer the reader to \cite{vampmain} for more details on the considered channel model and corresponding parameters. The simulation parameters are listed in Table \ref{table:sim-parameter}.

We also consider two line-of-sight (LOS) conditions. In the first case, we assume only the BS-RedRIS channel has a LOS component. In the second case, we assume both the BS-RedRIS and RedRIS-user channels have LOS components, but there is no LOS component in the BS-user channels. We gauge the performance of the proposed schemes against:


\begin{enumerate}
    \item A MIMO system assisted by one reflective RIS with optimized phase shifts and the MMSE precoding at the BS. 
    \item A MIMO system assisted by one RedRIS with an unoptimized (i.e., random) switching matrix and MMSE precoding at the BS.
\end{enumerate}

\subsection{Performance results with perfect CSI}
\subsubsection{Only BS-RIS channel has a LOS component}
This situation is commonly encountered in typical urban areas where the BS is usually situated at a distant location from the users and has no direct link to them due to blockage \cite{vampmain}. We consider a $32$-antenna BS that is serving $4$ users.

Fig. \ref{figg 5a} illustrates the achievable sum-rate versus the total transmit power $P$. It is observed that the optimized fully-connected RedRIS provides a higher sum-rate than the optimized reflective RIS for all transmit powers. {This is due to the ability of the RedRIS to direct different incident signals towards the intended users. The reflective RIS, not being able to independently steer all incoming paths, encounters conflicting requirements when optimizing the rate of multiple users. Its limited spatial selectivity is what leads it to inferior performance.} We also observe that the optimized partially-connected RedRIS with $N_p=20$ effective ports (selected via the reduction method developed in Section \ref{sec:reduction}) achieves almost the same sum-rate as the fully-connected RedRIS. The former does so while incurring much less control overhead and operational complexity.

Fig. \ref{figg 5b} shows a plot of the sum-rate against the total transmit power for the single-user scenario. {In this setting, the superior spatial selectivity offered by the RedRIS is irrelevant as the RIS has to only steer the incident signals to the sole user in the network. Actually, since all the available ports of the RedRIS must be active, a higher spatial selectivity is detrimental as power is directed in unnecessary avenues. The RedRIS also suffers from quantization errors as we have discretized angles by explicitely modelling the 2D DFT matrix in $\mathbf{U}$. Due to the aforementioned reasons, the reflective RIS achieves the highest sum-rate in fig. \ref{figg 5b}. The RedRIS with a two-port connection (selected by the reduction method developed in Section \ref{sec:reduction}) also attains a sum-rate close to the other schemes. While a smaller number of ports is preferred in the single-user scenerio the large quantization errors that result from having two angle bins alone degrade the performance of the two port RedRIS. Indeed, there should be an optimum number of ports that balance the tradeoff between spatial selectivity and angular resolution.} 

Fig. \ref{figg 5c} depicts the sum-rate against the number, $K$, of ports (in RedRIS) or reflection elements (in reflective RIS) at $30$ dBm transmit power. As expected, the sum-rate increases steadily with $K$. Although the sum-rate at small values of $K$ is similar for all optimized schemes, a significant gap is observed as $K$ increases. It is also seen that the optimized partially-connected RedRIS with only $N_p=20$ selected ports performs almost as good as the fully-connected RedRIS, with the advantage of less complexity and control overhead.

Fig. \ref{figg 5d} depicts the sum-rate against the number of users. {In line with theory, we see that the sum-rate increases with the number of single-antenna users up to $10$ users wherein both RIS configurations saturate. This is due to the fact that the channel matrix becomes rank deficient as we have only considered $10$ multi-path components in the BS-RedRIS channel. With a low number of users, enhanced spatial selectivity is paramount as the incident signals at the RIS must be directed along a few sparse paths. When there are a large number of users however, spatial selectivity becomes insignificant as users become densely populated and any redirected signal will strike a user. The reflective RIS performs better in this case due to its ability to choose the optimum path with infinite angular precision.}
\begin{figure*}[t]
\centering
\begin{subfigure}{.33\textwidth}
\includegraphics[width=\textwidth]{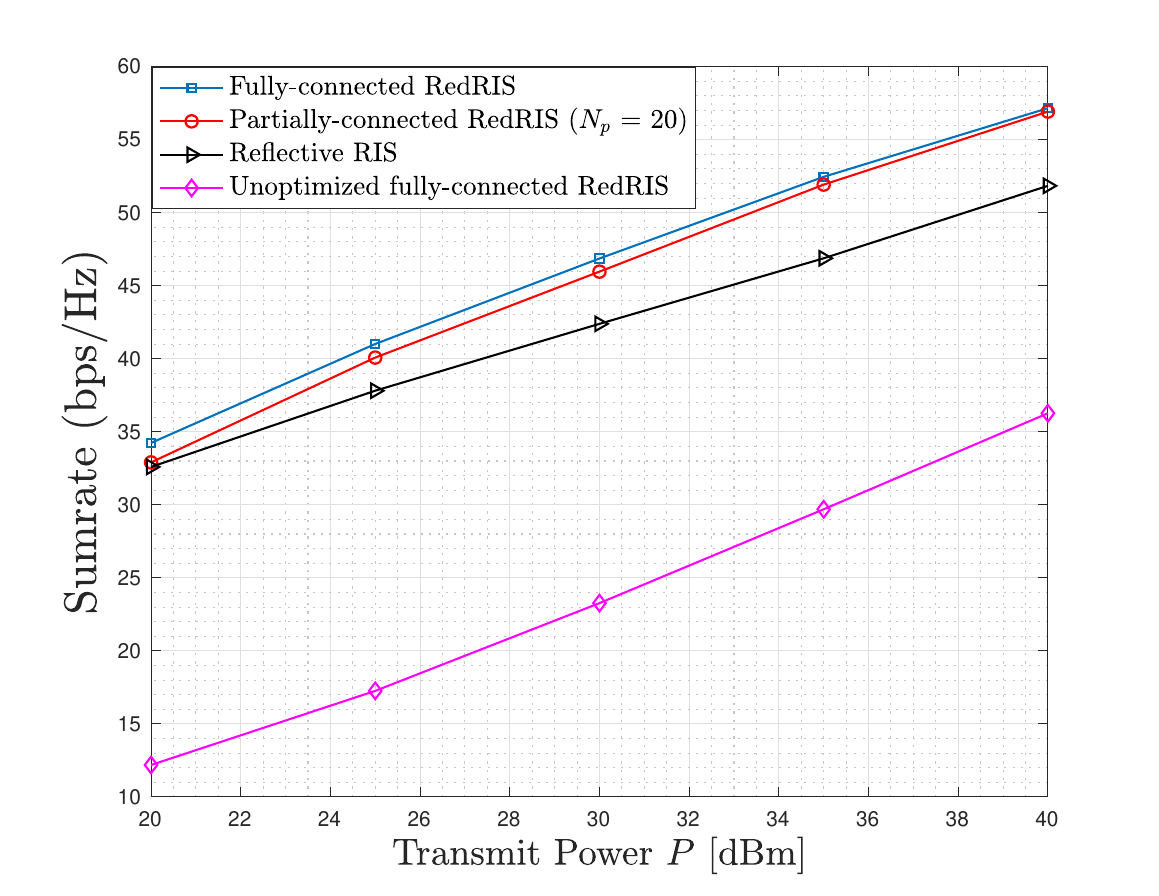}
\caption{\footnotesize $M=4, \ N=32$ and $K=256$.}
\label{figg 6a}
\end{subfigure}%
\hfill
\begin{subfigure}{0.33\textwidth}
\includegraphics[width = \textwidth]{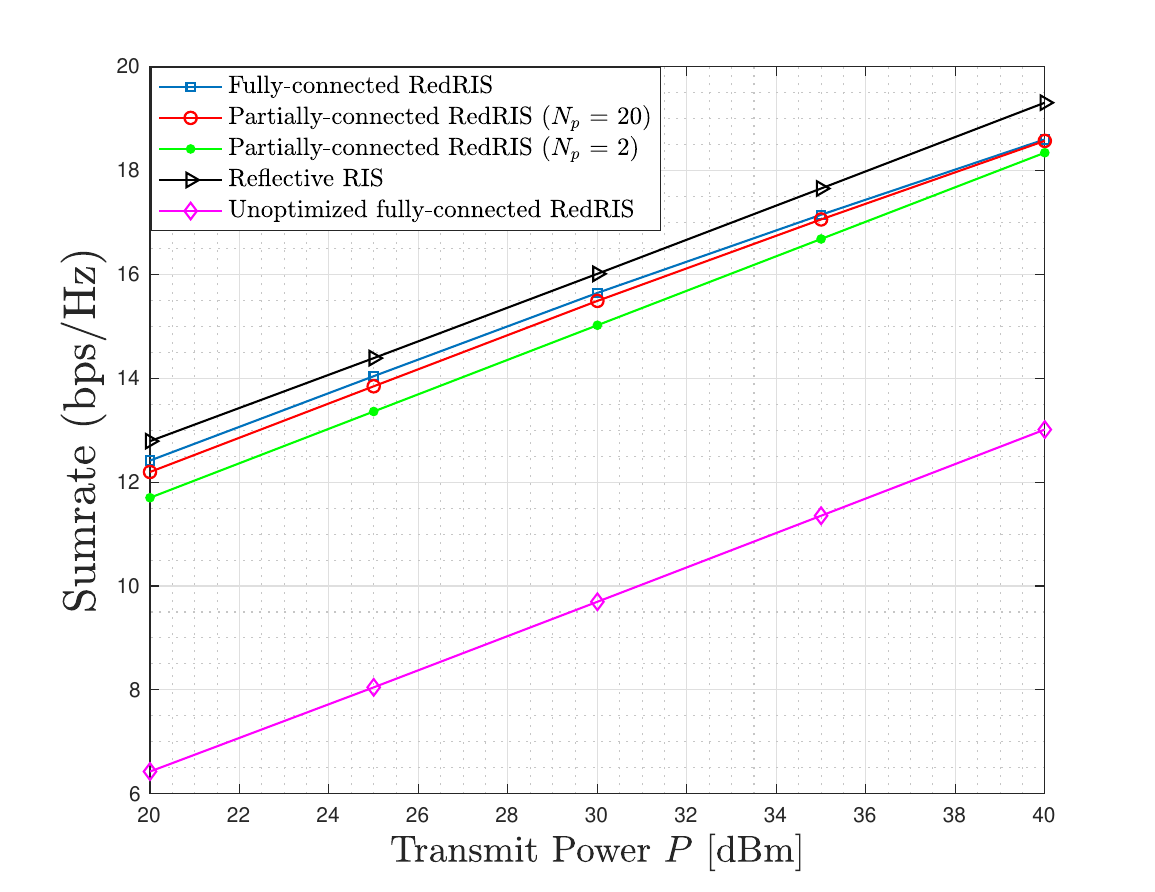}
\caption{\footnotesize $M=1, \ N=32$ and $K=256$.}
\label{figg 6b}
\end{subfigure}%
\hfill
\begin{subfigure}{.33\textwidth}
\includegraphics[width = \textwidth]{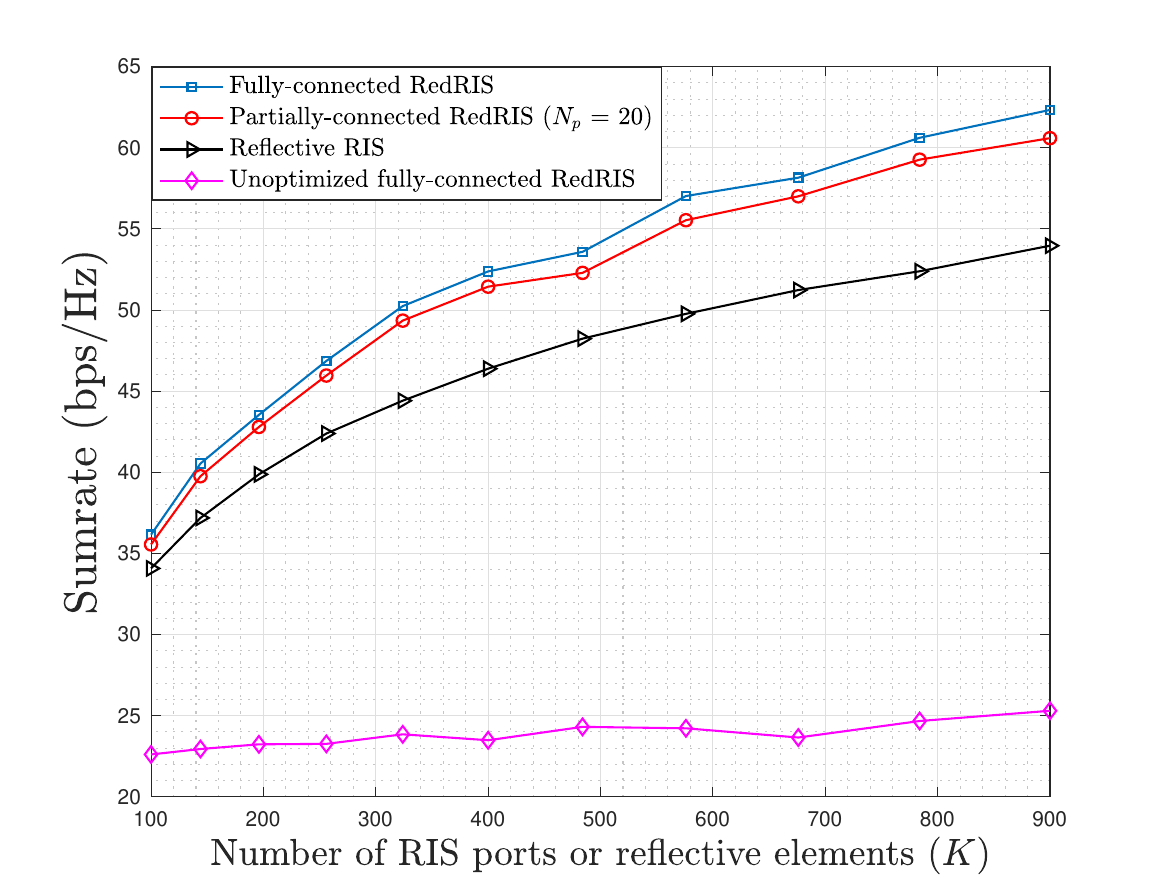}
\caption{\footnotesize $M=4, \ N=32$ and $P=30$ dBm.}
\label{figg 6c}
\end{subfigure}
\caption{(a) sum-rate versus transmit power (multi-user case), (b) sum-rate versus transmit power (single-user case), and (c) sum-rate versus the number of RIS ports or reflection elements (multi-user case). LOS components are present in both the BS-RIS channel and the RIS-users channels.}
\end{figure*}

\subsubsection{BS-RIS channel and RIS-user channels have LOS components}
Figs. \ref{figg 6a} and \ref{figg 6b} depict the sum-rate performance versus the transmit power for multi- and single-user cases, respectively. As expected, it is observed that the presence of LOS components in the RIS-users channels improves the sum-rate appreciably in both scenarios. Fig. \ref{figg 6c} shows the sum-rate against the number of ports (or reflection elements) in presence of a LOS component which is an upwardly shifted version of the graph for the NLOS scenario due to the improved sum-rate.
\begin{figure*}
\centering
\begin{subfigure}{.5\textwidth}
\includegraphics[width = \textwidth]{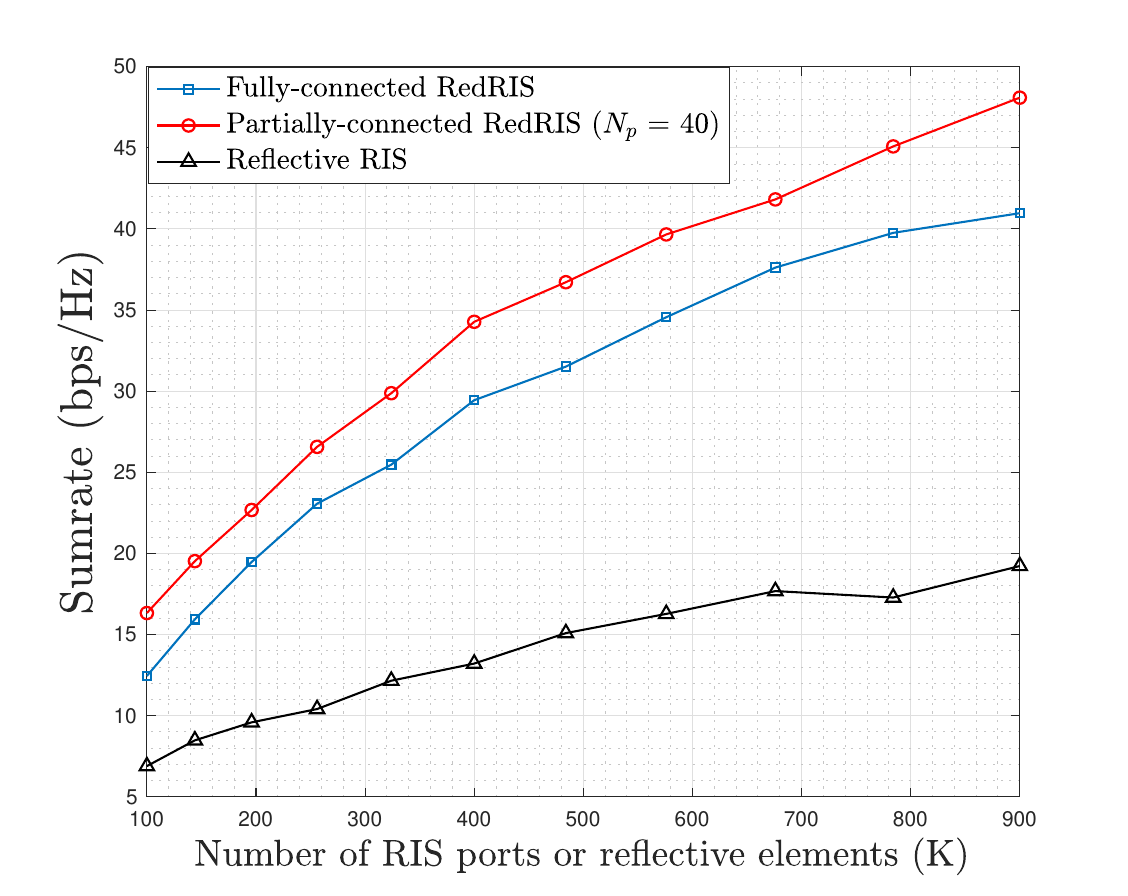}
\caption{\footnotesize $M=8, \ BS=8$ and $P=30$ dBm.}
\label{figg 8a}
\end{subfigure}%
\hfill
\begin{subfigure}{.5\textwidth}
\includegraphics[width = \textwidth]{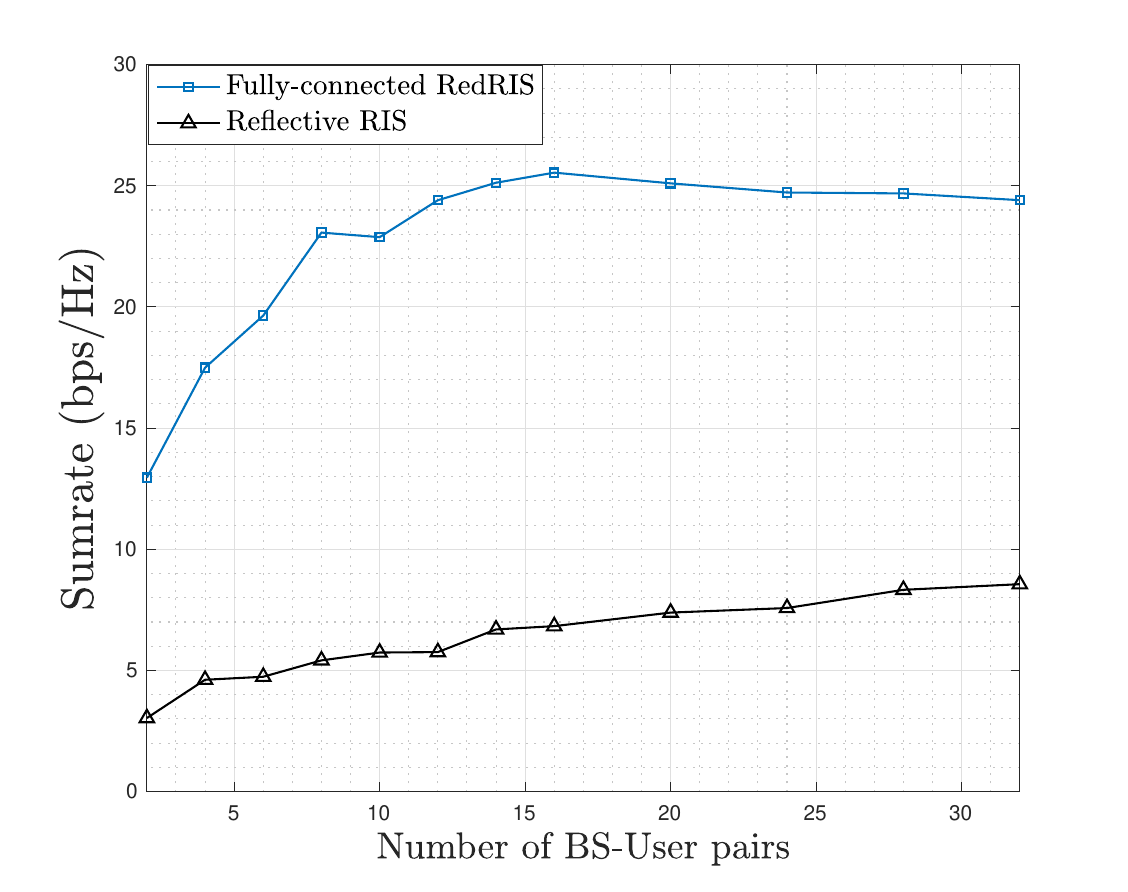}
\caption{\footnotesize $K=256$ and $P=30$ dBm.}
\label{figg 8b}
\end{subfigure}
\caption{(a) sum-rate versus the number of RIS ports or reflection elements, (b) sum-rate versus the number of BS-user pairs in multi-cell scenario. A LOS component is present in {each of the} BS-RIS channels only.}
\end{figure*}

\subsection{Performance in the multi-cell scenario}
We consider a cluster consisting of $8$ cells each of which being equipped with a single-antenna BS and each BS is serving one user at a time with the help of the RedRIS. We consider the same parameters as in the single-cell configuration we already mentioned in \textbf{Table \ref{table:sim-parameter}}. Fig. \ref{figg 8a} compares the achievable sum-rate of the RedRIS and the reflective RIS against the number, $K$, of RedRIS ports or reflective RIS elements. There, it is seen that {both the fully- and partially-connected RedRIS} significantly outperform the widely adopted reflective RIS in the multi-cell setup. {The sum-rate for the RedRIS configurations increase with the number of port elements due to the enhanced angular resolution. The partially-connected RedRIS, choosing the optimum $N_p$ out of $K$ ports, outperforms the fully-connected RedRIS by avoiding the higher interference that comes with a larger number of ports. Similar to the single-cell setup both RedRIS configurations outperform the reflective RIS.}
Fig. \ref{figg 8b} plots the sum-rate versus the number of BS-user pairs for the multi-cell setup where the sum-rate with RedRIS increases up to a certain number of BS-user pairs and then saturates. With the increase of the number of BS-user pairs, the inter-cell interference also increases which limits any further gain in the sum-rate. The flat-line in the curve corresponds to the interference-limited regime. This figure shows that the sum-rate with RedRIS saturates at a higher value than the reflective RIS.


\begin{figure}
\centering
\includegraphics[scale=0.4]{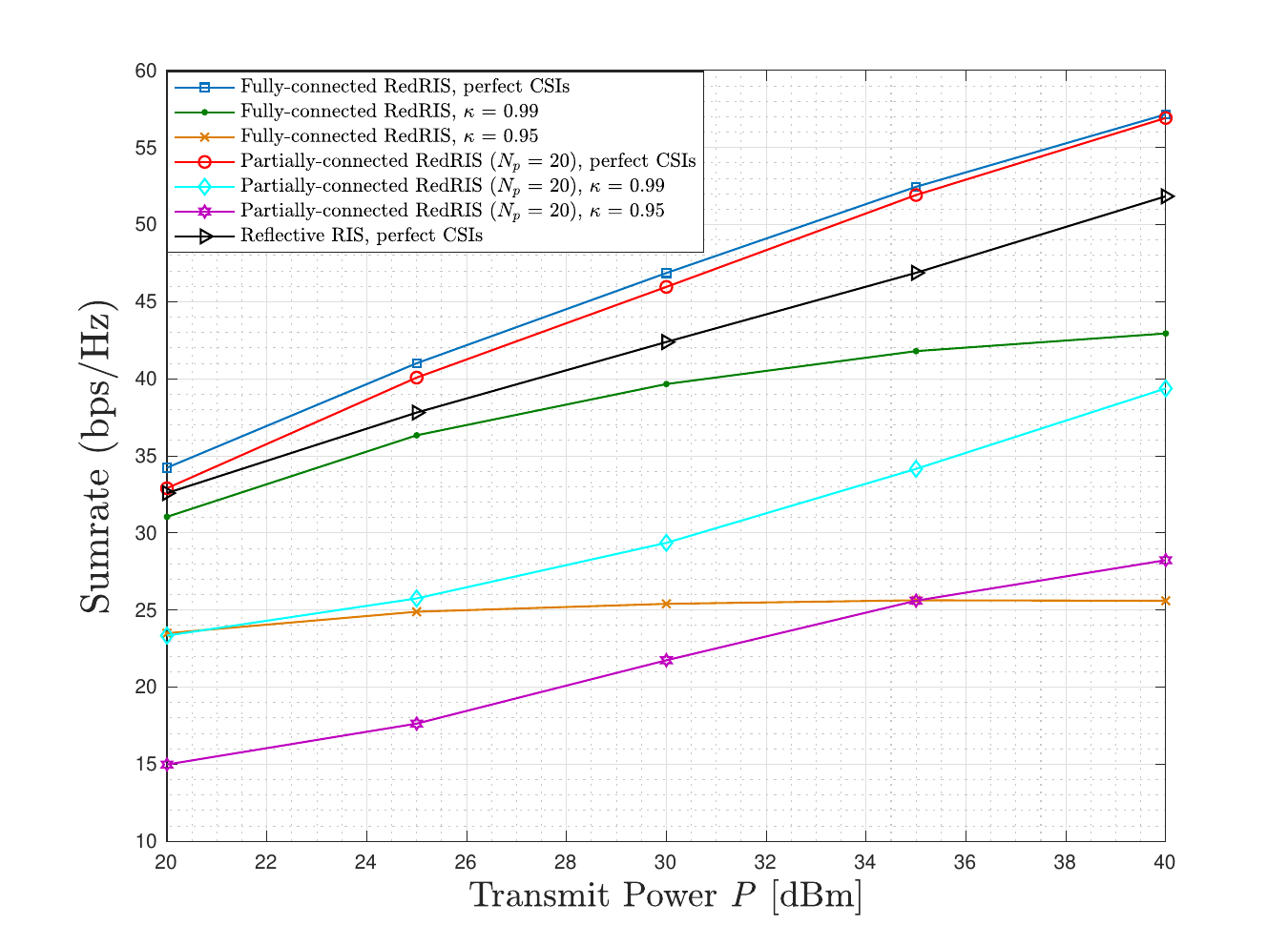}
\caption{Sum-rate versus transmit power under imperfect CSI ($M=4, \ N=32$ and $K=256$). LOS components are present in both the BS-RIS channel and the RIS-users channels.}
\label{figg 7}
\end{figure}
\subsection{Performance results with imperfect CSI}
In this section, we assess the performance of RedRIS-aided communication in presence of residual channel estimation errors. We refer the reader to \cite{vampmain} for the further details on how the latter are modeled and we exhibit the corresponding results in Fig. \ref{figg 7}. For comparison purposes, we also show the sum-rate curves under perfect CSI. In presence of small residual channel estimation errors, i.e., at $\kappa = 0.99$ and at low SNRs, {we see that the fully-connected RedRIS yields almost the same sum-rate as the perfect CSI cases. Actually at low SNR, the fully-connected RedRIS, with more channel error ($\kappa = 0.95$), performs just as well the partially-connected RedRIS with less error ($\kappa = 0.99$). Such performance degradation of the partially-connected RedRIS is caused by the induced errors in selecting the significant ports due to the imperfect CSI knowledge.
At high SNR, however, we see that the partially-connected RedRIS performs as well ($\kappa = 0.99$), or better ($\kappa = 0.95$) than the fully-connected RedRIS due to the exacerbated interference of the latter.}
\section{Conclusion}

\label{sec:conclusion}

In this paper, we {investigated the idea} of using a lens-type redirective surface (called RedRIS) in multi-user MIMO communication and developed a dedicated algorithm to optimize the {associated} port switching matrix. To that end, we first formulated a joint optimization problem to maximize the spectral efficiency of the users by minimizing the sum MMSE of all users' received symbols in the single-cell scenario. The joint optimization problem consists of two sub-optimization tasks: $i)$ the receive scaling factor and the precoding matrix at the BS were optimized in closed form using Lagrange optimization, while $ii)$ the switching matrix of the RedRIS ports was optimized iteratively using alternating optimization. We then extended the proposed framework to the multi-cell scenario with single-antenna base stations each of which serving a single-user. We then proposed two methods for reducing the number of effective port connections in the RedRIS so as to reduce the control overhead and computational complexity. The simulation results reveal that the proposed RedRIS-based solution, even with a reduced number of port connections, performs better than the widely-adopted reflective RIS solution. Moreover, the resilience of RedRIS against residual channel estimation errors was validated empirically. Possible extensions of this research work may include RedRIS channel estimation and extending the framework to multi-user cell-free MIMO.

\bibliographystyle{IEEEtran}
\bibliography{reference}

\begin{thebibliography}{10}
\providecommand{\url}[1]{#1}
\csname url@samestyle\endcsname
\providecommand{\newblock}{\relax}
\providecommand{\bibinfo}[2]{#2}
\providecommand{\BIBentrySTDinterwordspacing}{\spaceskip=0pt\relax}
\providecommand{\BIBentryALTinterwordstretchfactor}{4}
\providecommand{\BIBentryALTinterwordspacing}{\spaceskip=\fontdimen2\font plus
\BIBentryALTinterwordstretchfactor\fontdimen3\font minus \fontdimen4\font\relax}
\providecommand{\BIBforeignlanguage}[2]{{%
\expandafter\ifx\csname l@#1\endcsname\relax
\typeout{** WARNING: IEEEtran.bst: No hyphenation pattern has been}%
\typeout{** loaded for the language `#1'. Using the pattern for}%
\typeout{** the default language instead.}%
\else
\language=\csname l@#1\endcsname
\fi
#2}}
\providecommand{\BIBdecl}{\relax}
\BIBdecl

\bibitem{9086766}
M.~A. ElMossallamy, H.~Zhang, L.~Song, K.~G. Seddik, Z.~Han, and G.~Y. Li, ``Reconfigurable intelligent surfaces for wireless communications: Principles, challenges, and opportunities,'' \emph{IEEE Transactions on Cognitive Communications and Networking}, vol.~6, no.~3, pp. 990--1002, 2020.

\bibitem{9326394}
Q.~Wu, S.~Zhang, B.~Zheng, C.~You, and R.~Zhang, ``Intelligent reflecting surface-aided wireless communications: A tutorial,'' \emph{IEEE Transactions on Communications}, vol.~69, no.~5, pp. 3313--3351, 2021.

\bibitem{9690477}
Z.~Chen, B.~Ning, C.~Han, Z.~Tian, and S.~Li, ``Intelligent reflecting surface assisted terahertz communications toward 6g,'' \emph{IEEE Wireless Communications}, vol.~28, no.~6, pp. 110--117, 2021.

\bibitem{8387211}
I.~F. Akyildiz, C.~Han, and S.~Nie, ``Combating the distance problem in the millimeter wave and terahertz frequency bands,'' \emph{IEEE Communications Magazine}, vol.~56, no.~6, pp. 102--108, 2018.

\bibitem{9618776}
H.-J. Song and N.~Lee, ``Terahertz communications: Challenges in the next decade,'' \emph{IEEE Transactions on Terahertz Science and Technology}, vol.~12, no.~2, pp. 105--117, 2022.

\bibitem{8647621}
Q.~Wu and R.~Zhang, ``Intelligent reflecting surface enhanced wireless network: Joint active and passive beamforming design,'' in \emph{2018 IEEE Global Communications Conference (GLOBECOM)}, 2018, pp. 1--6.

\bibitem{8917871}
Y.-C. Liang, R.~Long, Q.~Zhang, J.~Chen, H.~V. Cheng, and H.~Guo, ``Large intelligent surface/antennas (lisa): Making reflective radios smart,'' \emph{Journal of Communications and Information Networks}, vol.~4, no.~2, pp. 40--50, 2019.

\bibitem{8811733}
Q.~Wu and R.~Zhang, ``Intelligent reflecting surface enhanced wireless network via joint active and passive beamforming,'' \emph{IEEE Transactions on Wireless Communications}, vol.~18, no.~11, pp. 5394--5409, 2019.

\bibitem{renzo2019smart}
M.~D. Renzo, M.~Debbah, D.-T. Phan-Huy, A.~Zappone, M.-S. Alouini, C.~Yuen, V.~Sciancalepore, G.~C. Alexandropoulos, J.~Hoydis, H.~Gacanin \emph{et~al.}, ``Smart radio environments empowered by reconfigurable ai meta-surfaces: An idea whose time has come,'' \emph{EURASIP Journal on Wireless Communications and Networking}, vol. 2019, no.~1, pp. 1--20, 2019.

\bibitem{9362274}
B.~Zheng, C.~You, and R.~Zhang, ``Double-irs assisted multi-user mimo: Cooperative passive beamforming design,'' \emph{IEEE Transactions on Wireless Communications}, vol.~20, no.~7, pp. 4513--4526, 2021.

\bibitem{9473585}
J.~Kim, S.~Hosseinalipour, T.~Kim, D.~J. Love, and C.~G. Brinton, ``Multi-irs-assisted multi-cell uplink mimo communications under imperfect csi: A deep reinforcement learning approach,'' in \emph{2021 IEEE International Conference on Communications Workshops (ICC Workshops)}, 2021, pp. 1--7.

\bibitem{8930608}
Q.~Wu and R.~Zhang, ``Beamforming optimization for wireless network aided by intelligent reflecting surface with discrete phase shifts,'' \emph{IEEE Transactions on Communications}, vol.~68, no.~3, pp. 1838--1851, 2020.

\bibitem{9198125}
M.-M. Zhao, Q.~Wu, M.-J. Zhao, and R.~Zhang, ``Intelligent reflecting surface enhanced wireless networks: Two-timescale beamforming optimization,'' \emph{IEEE Transactions on Wireless Communications}, vol.~20, no.~1, pp. 2--17, 2021.

\bibitem{8683145}
Q.~Wu and R.~Zhang, ``Beamforming optimization for intelligent reflecting surface with discrete phase shifts,'' in \emph{ICASSP 2019 - 2019 IEEE International Conference on Acoustics, Speech and Signal Processing (ICASSP)}, 2019, pp. 7830--7833.

\bibitem{9206080}
H.~Yang, Z.~Xiong, J.~Zhao, D.~Niyato, L.~Xiao, and Q.~Wu, ``Deep reinforcement learning-based intelligent reflecting surface for secure wireless communications,'' \emph{IEEE Transactions on Wireless Communications}, vol.~20, no.~1, pp. 375--388, 2021.

\bibitem{9115725}
S.~Abeywickrama, R.~Zhang, Q.~Wu, and C.~Yuen, ``Intelligent reflecting surface: Practical phase shift model and beamforming optimization,'' \emph{IEEE Transactions on Communications}, vol.~68, no.~9, pp. 5849--5863, 2020.

\bibitem{9226616}
P.~Wang, J.~Fang, X.~Yuan, Z.~Chen, and H.~Li, ``Intelligent reflecting surface-assisted millimeter wave communications: Joint active and passive precoding design,'' \emph{IEEE Transactions on Vehicular Technology}, vol.~69, no.~12, pp. 14\,960--14\,973, 2020.

\bibitem{8741198}
C.~Huang, A.~Zappone, G.~C. Alexandropoulos, M.~Debbah, and C.~Yuen, ``Reconfigurable intelligent surfaces for energy efficiency in wireless communication,'' \emph{IEEE Transactions on Wireless Communications}, vol.~18, no.~8, pp. 4157--4170, 2019.

\bibitem{8796365}
E.~Basar, M.~Di~Renzo, J.~De~Rosny, M.~Debbah, M.-S. Alouini, and R.~Zhang, ``Wireless communications through reconfigurable intelligent surfaces,'' \emph{IEEE Access}, vol.~7, pp. 116\,753--116\,773, 2019.

\bibitem{wu2019intelligent}
Q.~Wu and R.~Zhang, ``Intelligent reflecting surface enhanced wireless network via joint active and passive beamforming,'' \emph{IEEE transactions on wireless communications}, vol.~18, no.~11, pp. 5394--5409, 2019.

\bibitem{wu2019beamforming}
------, ``Beamforming optimization for wireless network aided by intelligent reflecting surface with discrete phase shifts,'' \emph{IEEE Transactions on Communications}, vol.~68, no.~3, pp. 1838--1851, 2019.

\bibitem{abeywickrama2020intelligent}
S.~Abeywickrama, R.~Zhang, Q.~Wu, and C.~Yuen, ``Intelligent reflecting surface: Practical phase shift model and beamforming optimization,'' \emph{IEEE Transactions on Communications}, vol.~68, no.~9, pp. 5849--5863, 2020.

\bibitem{pan2020multicell}
C.~Pan, H.~Ren, K.~Wang, W.~Xu, M.~Elkashlan, A.~Nallanathan, and L.~Hanzo, ``Multicell mimo communications relying on intelligent reflecting surfaces,'' \emph{IEEE Transactions on Wireless Communications}, vol.~19, no.~8, pp. 5218--5233, 2020.

\bibitem{vampmain}
H.~Ur~Rehman, F.~Bellili, A.~Mezghani, and E.~Hossain, ``Joint active and passive beamforming design for irs-assisted multi-user mimo systems: A vamp-based approach,'' \emph{IEEE Transactions on Communications}, vol.~69, no.~10, pp. 6734--6749, 2021.

\bibitem{wijekoon2024phase}
D.~Wijekoon, A.~Mezghani, and E.~Hossain, ``Phase shifter optimization in ris-aided mimo systems under multiple reflections,'' \emph{IEEE Transactions on Wireless Communications}, 2024.

\bibitem{liu2021reconfigurable}
Y.~Liu, X.~Liu, X.~Mu, T.~Hou, J.~Xu, M.~Di~Renzo, and N.~Al-Dhahir, ``Reconfigurable intelligent surfaces: Principles and opportunities,'' \emph{IEEE communications surveys \& tutorials}, vol.~23, no.~3, pp. 1546--1577, 2021.

\bibitem{shastri2023nonlocal}
K.~Shastri and F.~Monticone, ``Nonlocal flat optics,'' \emph{Nature Photonics}, vol.~17, no.~1, pp. 36--47, 2023.

\bibitem{asadchy2016perfect}
V.~S. Asadchy, M.~Albooyeh, S.~N. Tcvetkova, A.~D{\'\i}az-Rubio, Y.~Ra'di, and S.~A. Tretyakov, ``Perfect control of reflection and refraction using spatially dispersive metasurfaces,'' \emph{Physical Review B}, vol.~94, no.~7, p. 075142, 2016.

\bibitem{estakhri2016wave}
N.~M. Estakhri and A.~Alu, ``Wave-front transformation with gradient metasurfaces,'' \emph{Physical Review X}, vol.~6, no.~4, p. 041008, 2016.

\bibitem{mezghani2022reconfigurable}
A.~Mezghani, F.~Bellili, and E.~Hossain, ``Reconfigurable intelligent surfaces for quasi-passive mmwave and thz networks: Should they be reflective or redirective?'' in \emph{2022 56th Asilomar Conference on Signals, Systems, and Computers}.\hskip 1em plus 0.5em minus 0.4em\relax IEEE, 2022, pp. 1076--1080.

\bibitem{shen2021modeling}
S.~Shen, B.~Clerckx, and R.~Murch, ``Modeling and architecture design of reconfigurable intelligent surfaces using scattering parameter network analysis,'' \emph{IEEE Transactions on Wireless Communications}, vol.~21, no.~2, pp. 1229--1243, 2021.

\bibitem{li2022reconfigurable}
Q.~Li, M.~El-Hajjar, I.~Hemadeh, A.~Shojaeifard, A.~A. Mourad, B.~Clerckx, and L.~Hanzo, ``Reconfigurable intelligent surfaces relying on non-diagonal phase shift matrices,'' \emph{IEEE Transactions on Vehicular Technology}, vol.~71, no.~6, pp. 6367--6383, 2022.

\bibitem{nerini2023discrete}
M.~Nerini, S.~Shen, and B.~Clerckx, ``Discrete-value group and fully connected architectures for beyond diagonal reconfigurable intelligent surfaces,'' \emph{IEEE Transactions on Vehicular Technology}, 2023.

\bibitem{nerini2023closed}
------, ``Closed-form global optimization of beyond diagonal reconfigurable intelligent surfaces,'' \emph{IEEE Transactions on Wireless Communications}, 2023.

\bibitem{nerini2024beyond}
M.~Nerini, S.~Shen, H.~Li, and B.~Clerckx, ``Beyond diagonal reconfigurable intelligent surfaces utilizing graph theory: Modeling, architecture design, and optimization,'' \emph{IEEE Transactions on Wireless Communications}, 2024.

\bibitem{nerini2023pareto}
M.~Nerini and B.~Clerckx, ``Pareto frontier for the performance-complexity trade-off in beyond diagonal reconfigurable intelligent surfaces,'' \emph{IEEE Communications Letters}, 2023.

\bibitem{mezghani2022nonlocal}
A.~Mezghani, F.~Bellili, and E.~Hossain, ``Nonlocal reconfigurable intelligent surfaces for wireless communication: Modeling and physical layer aspects,'' \emph{arXiv preprint arXiv:2210.05928}, 2022.

\bibitem{ayoubi2022network}
R.~A. Ayoubi, M.~Mizmizi, D.~Tagliaferri, D.~De~Donno, and U.~Spagnolini, ``Network-controlled repeaters vs. reconfigurable intelligent surfaces for 6g mmw coverage extension,'' \emph{arXiv preprint arXiv:2211.08033}, 2022.

\bibitem{zhang2021learning}
Y.~Zhang and A.~Alkhateeb, ``Learning reflection beamforming codebooks for arbitrary ris and non-stationary channels,'' \emph{arXiv preprint arXiv:2109.14909}, 2021.

\bibitem{https://doi.org/10.48550/arxiv.2003.02538}
\BIBentryALTinterwordspacing
A.~Zappone, M.~Di~Renzo, F.~Shams, X.~Qian, and M.~Debbah, ``Overhead-aware design of reconfigurable intelligent surfaces in smart radio environments,'' 2020. [Online]. Available: \url{https://arxiv.org/abs/2003.02538}
\BIBentrySTDinterwordspacing

\bibitem{9039554}
Y.~Yang, B.~Zheng, S.~Zhang, and R.~Zhang, ``Intelligent reflecting surface meets ofdm: Protocol design and rate maximization,'' \emph{IEEE Transactions on Communications}, vol.~68, no.~7, pp. 4522--4535, 2020.

\bibitem{6648436}
S.~V. Hum and J.~Perruisseau-Carrier, ``Reconfigurable reflectarrays and array lenses for dynamic antenna beam control: A review,'' \emph{IEEE Transactions on Antennas and Propagation}, vol.~62, no.~1, pp. 183--198, 2014.

\bibitem{dimitrinl}
D.~P. Bertsekas, ``{Nonlinear programming},'' \emph{Journal of the Operational Research Society}, vol.~48, no.~3, pp. 334--334, 1997.

\bibitem{ghouse19932d}
M.~A. Ghouse, ``2d grid architectures for the dft and the 2d dft,'' \emph{Journal of VLSI signal processing systems for signal, image and video technology}, vol.~5, no.~1, pp. 57--74, 1993.

\bibitem{heathmimo}
R.~W. Heath~Jr. and A.~Lozano, \emph{{Foundations of MIMO Communication}}.\hskip 1em plus 0.5em minus 0.4em\relax Cambridge University Press, 2018.

\bibitem{8081330}
H.~Jedda, A.~Mezghani, A.~L. Swindlehurst, and J.~A. Nossek, ``Precoding under instantaneous per-antenna peak power constraint,'' in \emph{2017 25th European Signal Processing Conference (EUSIPCO)}, 2017, pp. 863--867.

\bibitem{lagrange}
M.~Joham, W.~Utschick, and J.~Nossek, ``Linear transmit processing in mimo communications systems,'' \emph{IEEE Transactions on Signal Processing}, vol.~53, no.~8, pp. 2700--2712, 2005.

\bibitem{hungarian}
H.~W. Kuhn, ``The hungarian method for the assignment problem,'' \emph{Naval research logistics quarterly}, vol.~2, no. 1-2, pp. 83--97, 1955.

\bibitem{jiang2022interference}
T.~Jiang and W.~Yu, ``Interference nulling using reconfigurable intelligent surface,'' \emph{IEEE Journal on Selected Areas in Communications}, vol.~40, no.~5, pp. 1392--1406, 2022.

\end{thebibliography}

\end{document}